\documentclass[%
 pra,
 reprint,
 onecolumn,
 superscriptaddress,
 noeprint,
]{revtex4-2}
\usepackage[]{standalone}
\usepackage{import}
\usepackage{amsmath}
\usepackage{amssymb}
\usepackage{amstext}
\usepackage{amsfonts}
\usepackage{mathrsfs}
\usepackage{cases}
\usepackage{graphicx}
\usepackage{mathtools}
\DeclarePairedDelimiter\ceil{\lceil}{\rceil}

\usepackage[table,xcdraw]{xcolor}
\usepackage{times}
\usepackage{hyperref}
\usepackage{todonotes}
\usepackage{graphicx}
\usepackage[T1]{fontenc}
\usepackage[utf8]{inputenc}
\usepackage{amsfonts}
\usepackage{bm}
\usepackage{physics}
\usepackage{amsthm}
\usepackage{enumitem}
\usepackage{multirow}
\usepackage{tabularx}
\usepackage{longtable}
\usepackage{float}
\usepackage{soul}

\usepackage{tikz}
\usetikzlibrary{external}
\usetikzlibrary{
  shapes.geometric,
  arrows,
  arrows.meta,
  positioning,
  backgrounds,
  fit,
  intersections,
}
\usepackage{qcircuit}
\usepackage{oplotsymbl}

\usepackage{hyperref}
\usepackage[capitalise]{cleveref}
\hypersetup{
 colorlinks=true,
 linkcolor=blue,
 citecolor=blue,
 filecolor=blue,
 urlcolor=blue,
 anchorcolor=blue,
 bookmarksopen=true,
 bookmarksopenlevel=1,
 pdfpagemode=UseOutlines,
 pdfstartview={XYZ null null 1},
 linktocpage=true,
}

\theoremstyle{plain}
\newtheorem{thm}{Theorem}

\newtheorem{lem}[thm]{Lemma}

\theoremstyle{definition}

\theoremstyle{remark}

\crefname{ansatz}{Ansatz.}{Ansatzes.}
\newcommand{\vertiii}[1]{{\left\vert\kern-0.25ex\left\vert\kern-0.25ex\left\vert #1 
    \right\vert\kern-0.25ex\right\vert\kern-0.25ex\right\vert}}

\newcommand{\ra}{\rangle}
\newcommand{\la}{\langle}
\newcommand{\ua}{\uparrow}
\newcommand{\da}{\downarrow}

\newcommand{\arccosh}{\operatorname{arccosh}}
\newcommand{\arcsinh}{\operatorname{arcsinh}}
\newcommand{\pdag}{{\phantom{\dagger}}}

\newcommand{\mycomment}[1]{}

\graphicspath{{.}{./texfigs/}}

\makeatletter
\def\blfootnote{\xdef\@thefnmark{}\@footnotetext}
\makeatother

\begin{document}

{\blfootnote{This manuscript has been authored by UT-Battelle, LLC, under Contract No.~DE-AC0500OR22725 with the U.S.~Department of Energy. The United States Government retains and the publisher, by accepting the article for publication, acknowledges that the United States Government retains a non-exclusive, paid-up, irrevocable, world-wide license to publish or reproduce the published form of this manuscript, or allow others to do so, for the United States Government purposes. The Department of Energy will provide public access to these results of federally sponsored research in accordance with the DOE Public Access Plan.}}

\title{Quantum Algorithms for Ground-State Preparation and Green's Function Calculation}

\author{Trevor Keen} 
\email{tkeen1@vols.utk.edu} 
\affiliation{Department of Physics and Astronomy, %
University of Tennessee, %
Knoxville, Tennessee 37996, USA}

\author{Eugene Dumitrescu}
\email{dumitrescuef@ornl.gov} 
\affiliation{Quantum Computational Science Group, %
%Computational Sciences and Engineering Division,
Oak Ridge National Laboratory, %
Oak Ridge, Tennessee 37831, USA}

\author{Yan Wang}
\email{wangy2@ornl.gov}
\affiliation{Quantum Computational Science Group, %
%Computational Sciences and Engineering Division,
Oak Ridge National Laboratory, %
Oak Ridge, Tennessee 37831, USA}

%\date{October 2020}
\date{\today}

\begin{abstract}
We propose quantum algorithms for projective ground-state preparation and calculations of the many-body Green's functions directly in frequency domain. The algorithms are based on the linear combination of unitary (LCU) operations and essentially only use quantum resources. To prepare the ground state, we construct the operator ${\exp}(-\tau \hat{H}^2)$ using Hubbard-Stratonovich transformation by LCU and apply it on an easy-to-prepare initial state. Our projective state preparation procedure saturates the near-optimal scaling, $\order*{\frac{1}{\gamma\Delta} \log \frac{1}{\gamma\eta}}$, of other currently known algorithms, in terms of the spectral-gap lower bound $\Delta$, the additive error $\eta$ in the state vector, and the overlap lower bound $\gamma$ between the initial state and the exact ground state. It is straightforward to combine our algorithm with spectral-gap amplification technique to achieve quadratically improved scaling $\order*{1/\sqrt{\Delta}}$ for ground-state preparation of frustration-free Hamiltonians, which we demonstrate with numerical results of the $q$-deformed XXZ chain.
To compute the Green's functions, including the single-particle and other response functions, we act on the prepared ground state with the retarded resolvent operator $R(\omega + i\Gamma; \hat{H})$ in the LCU form derived from the Fourier-Laplace integral transform (FIT). Our resolvent algorithm has $\order*{\frac{1}{\Gamma^2} \log\frac{1}{\Gamma\epsilon}}$ complexity scaling for the frequency resolution $\Gamma$ of the response functions and the targeted error $\epsilon$, while classical algorithms for FIT usually have polynomial scaling over the error $\epsilon$. To illustrate the complexity scaling of our algorithms, we provide numerical results for their application to the paradigmatic Fermi-Hubbard model on a one-dimensional lattice with different numbers of sites.
\end{abstract}

\maketitle

%\tableofcontents

\section{Introduction}
Quantum algorithms executed on fault tolerant quantum computers can potentially tackle notoriously challenging quantum many-body problems across the physical sciences for which efficient classical algorithms remain elusive. These quantum simulation algorithms typically involve three sequential tasks: (i) prepare the required input state, (ii) evolve the system via unitary quantum operations, and (iii) measure the quantities of interest. For a quantum many-body system governed by the Hamiltonian $\hat{H}$, these tasks pertain to solve the eigen-spectrum of $\hat{H}$ and prepare the corresponding eigenstates or implement its time-evolution operator $e^{-it\hat{H}}$. For the latter, i.e., implementing time dynamics of the system and measuring the desired observables, the quantum resource cost, including the size of quantum registry (i.e., the number of qubits) and the number of quantum operations, are in most cases well understood and algorithms with optimal cost are known~\cite{Berry2015, Low2019, Ge2019, Lin2020}. However, for the former, i.e., preparing ground and excited states or superpositions and distributions over these states, the quantum resource requirements are not as well understood. Nevertheless, it has been shown that in general, for any integer $k \geq 2$, the complexity of the general $k$-local Hamiltonian problem (of solving the ground state energy) is QMA-complete~\cite{QMA, 3QMA, kitaev02book}, where QMA stands for Quantum Merlin Arthur complexity class, the quantum analog of NP complexity class.

There exist a variety of quantum algorithms to prepare states of interest. Some methods, such as the variational quantum eigensolver (VQE)~\cite{VQE, Wecker2015VHA, Reiner2019}, leverage the variational principle with classical optimization schemes to prepare ground states. The prepared ground states are typically variational and these methods rely on an external classical feedback loop to drive optimization. One nonvariational method to prepare exact eigenstates of a system is quantum phase estimation (QPE)~\cite{Abrams1999QPE}, where the ground or excited eigenstates of the system are post-selected by measuring the phase of the state with respect to controlled applications of the time evolution operator. Another nonvariational method, adiabatic state preparation (ASP)~\cite{Albash_2018}, uses the adiabatic theorem to evolve a prepared, known ground state of some Hamiltonian to that of a target Hamiltonian of interest. Third, iterative methods~\cite{Ge2019, Kyriienko2020, Lin2020, Bespalova2021} are recently proposed, where some ground-state projecting operator (usually as a function of the Hamiltonian) is iteratively applied onto a trial state until the ground states are projected out. After sufficient applications the trial state can be guaranteed to be the ground state up to exponentially small error. And fourth, thermal state preparation may be performed using imaginary-time evolution, where in the long imaginary-time limit the prepared state is essentially the ground state~\cite{Chowdhury2017}. Analog thermalization techniques, for example, using an ancilla as a controllable dissipative cooling reservoir~\cite{Metcalf20}, have also been recently investigated. Finally, there are also combined methods, such as adiabatic evolution followed by QPE~\cite{Bauer16}, and in principle one may combine the above methods by applying them in series.

Each of these methods has its own advantages and disadvantages. VQE can often be implemented with no ancillary overhead and with a low circuit depth. However, one must ensure that the VQE loop converge to a global  instead of local minimum; for certain hard problems, searching global minimum in high-dimensional variation-parameter space leads to exponential computation cost. On the other hand, for the nonvariational methods of preparing the ground state, the scaling of the algorithm often depends on a polynomial of the inverse spectral gap of the system~\cite{Abrams1999QPE, Bauer16, Albash_2018, Ge2019, Lin2020}, which indicates preparing the ground state could be computationally expensive for systems with a vanishingly small spectral gap~\cite{Pearson2019}. In addition, nonvariational preparations also typically require additional quantum resources, making them prohibitive for near-term applications. 

Once the state of interest is prepared, to carry on with the next two tasks in solving many-body physics problems, relevant quantum operations can be implemented on a quantum computer by unitary operators (quantum gates) and, in case of nonunitary quantum operations, a linear combinations of unitaries (LCU) method~\cite{LCU} or qubitization~\cite{Low2019}. Implementing a nonunitary operator via LCU or qubitization incurs additional costs, from the ancilla qubits to the repetitions in simulations for increasing the success probability.

Green's functions, denoted by $G(t)$ in time domain and its Fourier transform $G(\omega)$ in frequency domain, are useful in describing the many-body systems with interesting collective quantum phenomena. The single-particle Green's function describes the propagation of the dressed single-particle state, while two-particle Green's function, i.e., linear response functions, such as spin and charge response functions, describe the propagation of collective excitations. Since the ground-state Green's function $G(\omega)$ is related to the matrix elements of the resolvent of the Hamiltonian $R(\omega) = (\omega - \hat{H})^{-1}$ in the perturbed ground state resulted from general $n$-body operators, the poles of $G(\omega)$ encode the eigenenergies of the quasiparticles and collective excitations. Many physical properties of the system in question can be determined from the single-particle and two-particle Green's functions. For example, the spectral density and local density of states (LDOS) of the system is given by the imaginary part of the retarded single-particle Green's function $\Im G^R(\omega)$. Various methods have been recently proposed for calculating Green's functions and other related quantities~\cite{Roggero19, Roggero20, Kosugi20, Kosugi20linear, Endo20, Chen21} using quantum computing devices. For example, the Gaussian integral transformation (GIT)~\cite{Roggero20} is used to compute the spectral density of the system by employing the qubitization technique, while statistical sampling and QPE is recently used to compute the single-particle and two-particle Green's functions of simple molecules~\cite{Kosugi20, Kosugi20linear} and similar approach is applied to compute other dynamic linear response functions~\cite{Roggero19}. For near-term applications, Refs.~\onlinecite{Endo20, Chen21} utilize variational quantum simulations to calculate the Green's function. In all these methods, the ground state is either given by an oracle or obtained with variational algorithms.

In this work we propose algorithms for preparing ground states and computing Green's functions based on integral transformations of the ground-state projecting operator $f(\hat{H}) = e^{-\frac{1}{2}t^2 \hat{H}^2}$ and the resolvent operator $R(\omega)$, respectively. For the ground-state preparation, the Hubbard-Stratonovich integral transformation~\cite{HubbardHS1959, Stratonovich1957} is applied to $f(\hat{H})$ and implemented by LCU. Our projective method saturates the optimal scaling of previous methods~\cite{Ge2019, Lin2020} with an lower requirement on the precision of the ground-state energy than Ge~\textit{et al.}~\cite{Ge2019}. Our second result is a quantum algorithm for implementing the resolvent operator to compute single-particle Green's function. This algorithm is based on a Fourier-Laplace integral transform (FIT) used in conjunction with the LCU, and it essentially only uses quantum resource, in contrast to variational~\cite{Endo20, Chen21} and QPE~\cite{Roggero19, Kosugi20, Kosugi20linear} methods. Our method is also the first quantum algorithm to nonvariationally prepare the ground state and compute the Green's function directly in the frequency domain.

Our work is organized as follows. In \cref{sec:methods} we formulate the algorithm for ground-state preparation and compare the resulting scalings against previous results~\cite{Ge2019, Lin2020}. We then give the algorithm for constructing the resolvent operator and use it to calculate the single-particle Green's functions. In \cref{sec:FFqXXZ}, we apply spectral gap amplification~\cite{Somma2013, Chowdhury2017} in conjunction with our ground-state preparation procedure to the frustration-free $q$-deformed XXZ chain to show a quadratic speedup in terms of the spectral gap. In \cref{sec:hubbard} we apply our algorithms to the paradigmatic Fermi-Hubbard model on a one-dimensional lattice with various numbers of sites. We prepare the ground state and compute the Green's functions and the local density of states, while presenting the empirical scalings of our algorithms. In \cref{sec:conclusion}, we discuss our results and some future work.

\section{Methods} \label{sec:methods}

\subsection{Projection based ground-state preparation} \label{sec:GSP}
To improve the Harrow, Hassidim, and Lloyd (HHL) algorithm~\cite{Harrow2009} for the quantum linear system problem (QLSP), Childs \textit{et al.}~\cite{Childs2017} introduced the following Gaussian integral representation for the inverse of a nonsingular operator $\hat{H}$
\begin{align} \label{eq:Hinv}
     \hat{H}^{-1}
  &= \frac{i}{\sqrt{2\pi}} \int_0^\infty dy \int_{-\infty}^\infty dz\,
     z e^{-\frac{1}{2}z^2} e^{-iyz \hat{H}}.
\end{align}
Discretized in a double Riemann sum, \cref{eq:Hinv} can be approximately implemented by LCU. Kyriienko~\cite{Kyriienko2020} generalized this equation (replacing the $z$ factor in the integrand with $zy^{K-1}$) to represent the operator $\hat{H}^{-K}$ for a positive integer $K$ and used this operator to project out ground states of suitable Hamiltonians by acting with it on an initial state with a finite overlap with the true ground state. This method applies to Hamiltonians with a ground energy having the smallest magnitude among all eigenenergies, such as Hamiltonians with a positive spectrum. For Hamiltonians with a nonpositive spectrum, Bespalova and Kyriienko~\cite{Bespalova2021} proposed using instead the operator $\hat{H}^{K}$ to project out the ground state. $\hat{H}^{K}$ can be expressed by a differential representation involving the time-evolution operator and the finite difference approximation to the derivatives can be evaluated by LCU~\cite{Seki2021} or quantum-classical hybrid algorithm~\cite{Bespalova2021, Guzman2021}.

However, using the double integral representation to construct the operator $\hat{H}^{-K}$ is an overkill for ground-state projection. This can be seen as follows. Dropping the $z$ factor from the integrand (this factor is unnecessary for a positive-definite $\hat{H}$), we find that the Gaussian $z$-integral in \cref{eq:Hinv} is a Fourier transform of the Gaussian $e^{-\frac{1}{2}z^2}$ and thus, after being performed analytically, gives another Gaussian $e^{-\frac{1}{2}y^2\hat{H}^2}$. This operator can already be used for projecting out the ground state of Hamiltonians with nonnegative spectrum, without the need for a second integral transformation to obtain $\hat{H}^{-K}$.

Our ground-state preparation procedure is based on the above idea. Apply the Hubbard-Stratonovich transformation~\cite{Stratonovich1957, HubbardHS1959} to the ground-state projecting operator
\begin{align} \label{eq:fullHS}
     f(\hat{H}) = e^{-\frac{1}{2}t^2 \hat{H}^2}
  &= \frac{1}{\sqrt{2\pi}} \int_{-\infty}^\infty dz\,
     e^{ -\frac{1}{2}z^2} e^{- i z t\hat{H} },
\end{align}
where variable $y$ in \cref{eq:Hinv} is changed to $t$ as analogous to the imaginary time $\tau$ in the imaginary-time evolution operator $e^{-\tau \hat{H}}$. After discretizing and truncating the integral in \cref{eq:fullHS}, we implement by LCU the following approximate operator
\begin{align}
    \label{eq:discretizedHS}
     h(\hat{H}) 
  &= \frac{1}{\sqrt{2\pi}} \sum_{k=-N_z}^{N_z}\Delta_z
     e^{ -\frac{1}{2}z_{k}^2} e^{-i z_{k} t\hat{H} }
  \equiv \sum_{k=-N_z}^{N_z} \alpha_k U_k,
  \quad (z_k = k\Delta_z).
\end{align}
Since $\hat{H}^2$ instead of $\hat{H}$ is used in $f(\hat{H})$, it is the eigenstate corresponding to the eigenvalue with the smallest magnitude that is projected out. However, it is $\hat{H}$ that appears in the time-evolution unitary operators of LCU, as a result of the Hubbard-Stratonovich transformation.

Suppose that the ground energy $\lambda_0$ is only known within a given precision $\delta_0$ to a given parameter $\bar{\lambda}_0$, i.e., $|\lambda_0 - \bar{\lambda}_0|\leq \delta_0$, and the precision satisfies $2\delta_0 < \Delta \leq \Delta_s$, where $\Delta$ is a given lower bound of the exact spectral gap $\Delta_s$ of the system. Then, the spectrum (the eigenvalue set) of $\hat{H}$ can be shifted to nonnegative domain by adding to $\hat{H}$ a constant $E_c$ that satisfies $-\bar{\lambda}_0 + \delta_0 \leq E_c < -\bar{\lambda}_0 + \Delta/2$. The ground state of the original $\hat{H}$ now corresponds to the eigenenergy with the smallest magnitude. In addition, the shifted ground energy $\lambda_0 + E_c$ and the first excited-state energy $\lambda_1 + E_c = \lambda_0 + \Delta_s + E_c$ satisfy $0 \leq \lambda_0 + E_c < \Delta \leq \Delta_s \leq \lambda_1 + E_c$. Next, assume that the spectrum is bounded from above so the domain of spectrum can be scaled to unity. Under these assumptions, we only consider a Hamiltonian $\hat{H}$ that has been shifted and normalized, i.e., the spectrum $\sigma(\hat{H})$ is a subset of the domain $[0, 1]$.

For ground state preparation, using a sufficiently large $t = \order{\frac{1}{\Delta} \sqrt{\log \frac{1}{\gamma \eta}}}$ in \cref{eq:fullHS} guarantees that the resulted state $\ket{\psi} = e^{-\frac{1}{2} t^2 \hat{H}^2} \ket{\psi_0}$ is within fidelity $1 - \epsilon$ to the ground state $\ket{\lambda_0}$. Here, $\ket{\psi_0}$ is a chosen trial state that has a nonzero overlap with the true ground state. This result is summarized in \cref{lem:tbound} and the proof is given in \cref{app:tboundProof}.

\begin{lem} \label{lem:tbound}
Consider a Hamiltonian $\hat{H}$ with the spectrum $\sigma(\hat{H}) \subseteq [0, 1]$, the spectral gap $\Delta_s \geq \Delta > 0$, the ground state $\ket{\lambda_0}$, and the ground energy $\lambda_0 \geq 0$. Given the operator $f(\hat{H}) = e^{-\frac{1}{2} t^2 \hat{H}^2}$ and a trial state $\ket{\psi_0}$ with a fidelity $\abs{\ip{\lambda_0}{\psi_0}} \geq \gamma > 0$ to the ground state, the normalized state $\ket{\psi} = f(\hat{H}) \ket{\psi_0} / \norm*{f(\hat{H}) \ket{\psi_0}}$ is within fidelity $1 - \epsilon$ to the ground state, that is, 
\begin{align*} \textstyle
       1 - \abs{\ip{\lambda_0}{\psi}} 
  \leq \frac{1}{2}\norm{\ket{\psi} - \ket{\lambda_0}} 
  \leq \frac{\eta^2}{2} \equiv \epsilon,
\end{align*}
if
\begin{align*} \textstyle
  t \geq \frac{1}{\Delta}\sqrt{2\log\frac{1}{\gamma\eta}} =
\order{\frac{1}{\Delta}\sqrt{\log\frac{1}{\gamma\eta}}}.
\end{align*}
\end{lem}

When the operator $f(\hat{H})$ in \cref{lem:tbound} is implemented in the form $h(\hat{H})$ operator in \cref{eq:discretizedHS} by LCU, our result on the complexity and quantum resource cost for preparing ground state with $h(\hat{H})$ operator is summarized in \cref{thm:lcuHST} and the proof is given in \cref{app:thm1}.

\begin{thm} \label{thm:lcuHST}
Consider a Hamiltonian $\hat{H}$ with the spectrum $\sigma(\hat{H}) \subseteq [0, 1]$, the spectral gap $\Delta_s \geq \Delta > 0$, the ground state $\ket{\lambda_0}$, and the ground energy $\lambda_0 \geq 0$. Given the LCU operator $h(\hat{H}) = \sum_{k=-N_z}^{N_z} \alpha_k U_k$, where $\alpha_k = \frac{\Delta_z}{\sqrt{2\pi}} e^{-\frac{1}{2}z_k^2}$, $U_k = e^{-i z_k t \hat{H}}$, and $z_k = k \Delta_z$, and a trial state $\ket{\psi_0}$ with a fidelity $\abs{\ip{\lambda_0}{\psi_0}} \geq \gamma > 0$ to the ground state, the normalized state $\ket{\psi} = h(\hat{H}) \ket{\psi_0} / \norm*{h(\hat{H}) \ket{\psi_0}}$ is within fidelity $1 - \epsilon$ to the ground state, that is, 
\begin{align*} \textstyle
       1 - \abs{\ip{\lambda_0}{\psi}} 
  \leq \frac{1}{2}\norm{\ket{\psi} - \ket{\lambda_0}}
  \leq \frac{(5\eta)^2}{2} \equiv \epsilon,
\end{align*}
if
\begin{align*}
  t &\textstyle \geq \frac{1}{\Delta}\sqrt{2\log\frac{1}{\gamma\eta}}
  = \order{\frac{1}{\Delta}\sqrt{\log\frac{1}{\gamma\eta}}},
\quad
  \lambda_0 \leq \frac{1}{t}
  =  \order{\Delta / \sqrt{\log\frac{1}{\gamma\eta}}},
\displaybreak[1] \\
  z_c &\textstyle = N_z \Delta_z
  \geq \sqrt{2\log\frac{2}{\gamma\eta}}
  = \order{\sqrt{\log\frac{1}{\gamma\eta}}},
\displaybreak[1] \\  
  \Delta_z &\textstyle = \frac{2\pi}{z_c + t}
  = \order{\Delta / \sqrt{\log\frac{1}{\gamma\eta}}},
\quad
  N_z = \ceil*{\frac{z_c}{\Delta_z}}
  = \order{\frac{1}{\Delta} \log \frac{1}{\gamma\eta}}.
\end{align*}
Implementing the LCU operator $h(\hat{H})$ requires $\order{\frac{\alpha}{\gamma \Delta} \log \frac{1}{\gamma \eta}}$ quantum queries to a time-evolution oracle for Hamiltonian $\hat{H}$ and $\order{\log\frac{1}{\Delta} + \log\log\frac{1}{\gamma\eta}}$ ancilla qubits using the standard formulation of LCU~\cite{LCU}. Here, $\alpha = \sum_{k=-N_z}^{N_z} |\alpha_k| = \order{1}$ is the $L_1$ norm of the coefficients in the LCU.
\end{thm}

Our projective ground-state preparation algorithm can be compared with the aforementioned projective or iterative methods~\cite{Lin2020, Ge2019, Kyriienko2020, Bespalova2021}. Notably, in terms of cost and efficiency, it has significant advantage over the inverse power iterative method in Ref.~\onlinecite{Kyriienko2020} using the operator $\hat{H}^{-K}$ by a double-integral representation. In \cref{tab:scaling}, we compare the asymptotic complexities of various ground-state preparation algorithms in terms of the time-evolution query complexities, the required number of ancilla qubits, and the required precision to the \textit{a priori} known ground energy. Comparing with the algorithm by Ge \textit{et al.}~\cite{Ge2019} using the operator $\cos^M(\hat{H})$ ($M$ is a sufficiently large integer), we find very similar results. After tightening a few bounds they used to derive the original scaling and cost (see \cref{app:ge_bounds}), we find identical asymptotic scalings. Our numerical results on XXZ chain and Fermi-Hubbard chain confirm this, and our algorithm shows small advantage in some of the examples of Hamiltonian models. Both algorithms saturate the near-optimal scaling proved by Lin and Tong~\cite{Lin2020}. Refs.~\onlinecite{Ge2019, Lin2020} also extended their respective algorithms to prepare ground state with unknown ground energy. By combining a quantum search subroutine with the ground-state projection operator, our algorithm can also be be applied in such cases by first estimating the ground energy to the required the precision, following similar strategy given in Ref.~\cite{Ge2019}.

\subsubsection{Improved scaling with spectral gap amplification} \label{sec:GapAmp}
The Hubbard-Stratonovic transformation was also previously applied to prepare thermal Gibbs state~\cite{Chowdhury2017}, where a spectral-gap amplified Hamiltonian $\hat{H}_r$ satisfying $\hat{H}_r^2 (\ket{\mathbf{0}}_{a_r} \otimes \ket{\psi}) = \ket{\mathbf{0}}_{a_r} \otimes (\hat{H} \ket{\psi})$ is constructed using the spectral-gap amplification technique~\cite{Chowdhury2017, Somma2013}, and the Hubbard-Stratonovic transformation is then applied to the thermal density operator as follows.
\begin{align}
     \ket{\mathbf{0}}_{a_r} \otimes
     \qty(e^{-\frac{1}{2} t^2 \hat{H}} \ket{\psi})
  &= \qty(e^{-\frac{1}{2} t^2 \hat{H}_r^2})
     \qty(\ket{\mathbf{0}}_{a_r} \otimes \ket{\psi}) \notag \\
  &= \qty(\frac{1}{\sqrt{2\pi}} \int_{-\infty}^{\infty} dz\,
     e^{-\frac{1}{2}z^2} e^{-izt\hat{H}_r})
     \qty(\ket{\mathbf{0}}_{a_r} \otimes \ket{\psi}),
     \label{eq:HSGapAmp}
\end{align}
where $\ket{\mathbf{0}}_{a_r}$ denotes the ancilla qubits for constructing the spectral-gap amplified Hamiltonian $\hat{H}_r$ whose action is equivalent to that of the square-root of $\hat{H}$. As mentioned before, in the long-imaginary time limit ($t\to \infty$), the thermal Gibbs state essentially becomes the pure ground state, so the algorithm from Ref.~\onlinecite{Chowdhury2017} can be directly applied to prepare the ground state. However, the operator $e^{-\frac{1}{2} t^2 \hat{H}}$ alone can be used as a ground-state projection operator without applying the full thermal state preparation algorithm that carries a large amount of unnecessary cost. For frustration-free (FF) Hamiltonians, our LCU algorithm and \cref{thm:lcuHST} are also applicable to the Hubbard-Stratonovic transformation in \cref{eq:HSGapAmp} involving Hamiltonian $\hat{H}_r$. Since the spectral gap of $\hat{H}_r$ is essentially $\sqrt{\Delta}$, the query complexity of our ground-state preparation algorithm combined with spectral gap amplification is reduced from $\order*{1/\Delta}$ to $\order*{1/\sqrt{\Delta}}$, except for the cost arising from using ancilla qubits $\ket{\mathbf{0}}_{a_r}$ to construct (i.e., block-encode) $\hat{H}_r$ and using time-evolution oracle of $\hat{H}_r$ instead of $\hat{H}$. Note that the quadratic speedup is only possible for FF Hamiltonians. In \cref{sec:FFqXXZ} we discuss more details on the ground-state preparation of FF Hamiltonians and illustrate quadratic speedup in preparing the ground states of the FF $q$-deformed XXZ chain. For \emph{nearly} FF Hamiltonians, Ref.~\cite{Thibodeau2021} proposed a completely different algorithm giving a scaling \emph{between} $\order*{1/\Delta}$ and $\order*{1/\sqrt{\Delta}}$.

\begin{table}[]
\renewcommand{\arraystretch}{4}
\centering
%\resizebox{\textwidth}{!}{%
\begin{tabular}{|c|c|c|c|}
\hline %\hline
%\rowcolor[HTML]{EFEFEF}[\tabcolsep]
   \textbf{Algorithm}
  &\parbox[c]{4cm}{\textbf{Query Complexity}\\
    ($\alpha$, $\Delta$, $\gamma$, $\eta$)}
  &\parbox[c]{4cm}{\textbf{Ancilla Qubits}\\
    ($\Delta$, $\gamma$, $\eta$)}
  &\parbox[c]{4cm}{\textbf{Required Ground\\Energy Precision}}
\\ \hline
%\cellcolor[HTML]{EFEFEF}
  This Work
  &$\displaystyle \order{\frac{\alpha}{\gamma\Delta} \log\frac{1}{\gamma \eta}}$
  &$\displaystyle \order{\log\frac{1}{\Delta} + \log\log\frac{1}{\gamma \eta}}$
  &$\displaystyle \order{\flatfrac{\Delta}{\log^{1/2}\frac{1}{\gamma \eta}}}$
\\ \hline
  Ge \textit{et al.}~\cite{Ge2019}
  &$\displaystyle \order{ \frac{\alpha}{\gamma\Delta} \log^{3/2}\frac{1}{\gamma \eta}}$
  &$\displaystyle \order{\log\frac{1}{\Delta} + \log\log\frac{1}{\gamma \eta}}$
  &$\displaystyle \order{\flatfrac{\Delta}{\log\frac{1}{\gamma \eta}}}$
\\ \hline
  Improved Ge \textit{et al.}~\cite{Ge2019}
  &$\displaystyle \order{ \frac{\alpha}{\gamma\Delta} \log\frac{1}{\gamma \eta}}$
  &$\displaystyle \order{\log\frac{1}{\Delta} + \log\log\frac{1}{\gamma \eta}}$
  &$\displaystyle \order{\flatfrac{\Delta}{\log^{1/2}\frac{1}{\gamma \eta}}}$
\\ \hline %\rowcolor[HTML]{EFEFEF}
%\cellcolor[HTML]{EFEFEF}
  Lin and Tong~\cite{Lin2020}
  &$\displaystyle \order{\frac{\alpha}{\gamma\Delta} \log\frac{1}{\gamma \eta}}$
  &$\displaystyle \order{1}$
  &\parbox[c]{4cm}{$\Delta \leq \Delta_{\text{s}}$\\
    $ \dfrac{\Delta}{2}\leq \mu - \lambda_0 \leq 
     \Delta_{\text{s}} - \dfrac{\Delta}{2}$}
\\ \hline %\hline
\end{tabular}
%}
\caption{Comparison of the complexity scaling for ground-state preparation with known ground energy. Here, $\alpha$ is the $L_1$ norm of the coefficients of the LCU, except in the case of Lin and Tong, where $\alpha $ refers to the $(\alpha, m, 0)$ block encoding of the Hamiltonian, $\Delta_s$ is the exact spectral gap of a given Hamiltonian, $\Delta$ is a given spectral-gap lower bound, $\gamma$ is the lower bound of the overlap between the initial trial state and true ground state, and $\eta$ is the additive error in the state vector.}
  \label{tab:scaling}
\end{table}
%\end{landscape}

\subsection{Green's functions for many-body systems}
In an interacting many-body quantum system, single-particle modes $\ket{l}$ label the quantum states of an individual particle. Denote $\ket{l} \equiv \hat{c}_l^\dagger\ket{}$, where $\hat{c}_l^\dagger$ ($\hat{c}_l^\pdag$) creates (annihilates) a particle of the mode $l$ and $\ket{}$ is the zero-particle vacuum state. $l$ can be a set of combined indices necessary to characterize the quantum state. For spin-$\frac{1}{2}$ fermions on a lattice, $\ket{l} = \ket{j\sigma} = \hat{c}_{j\sigma}^\dagger\ket{}$ and these modes are labeled by the lattice site index $j\in \{1,2,\dots,L\}$ ($L$ is the number of lattice sites) and the spin index $\sigma\in \{\ua, \da\}$. For condensed matter systems, the lattice is usually the Bravais lattice with symmetries from certain space group that includes translation group as a subgroup; for atoms and molecules, the lattice usually assumes symmetries from certain point group and these modes are often referred as spin-orbitals.

The single-particle Green's function describes the response of an interacting many-body system to the perturbation of injecting and later removing one particle. Specifically, for an $N$-particle system, at zero temperature $T=0$, the retarded double-time Green's function in time domain is given by
%\begin{subequations}
\begin{align}
  \label{eq:GFt}
    G_{j\sigma, j'\sigma'}^R(t,t')
 &= -i\theta(t-t')\ev{[\hat{c}_{j\sigma}^\pdag (t),
 \hat{c}_{j'\sigma'}^\dagger (t')]_{+}}{\psi_0^N}, 
\end{align}
%\end{subequations}
where $\ket{\psi_0^N}$ is the ground state of the $N$-particle system, $[\hat{A}, \hat{B}]_{\zeta} = \hat{A}\hat{B} + \zeta \hat{B} \hat{A}$, and the creation and annihilation operators in Heisenberg picture are defined as
\begin{align}
    \hat{c}_{j\sigma}^\pdag (t)
  = e^{it(\hat{H} - \mu \hat{N})} \hat{c}_{j\sigma}^\pdag e^{-it(\hat{H} - \mu \hat{N})}, \quad
    \hat{c}_{j'\sigma'}^\dagger (t')
  = e^{it'(\hat{H} - \mu \hat{N})} \hat{c}_{j'\sigma'}^\dagger e^{-it'(\hat{H} - \mu \hat{N})}.
\end{align}
Here, we include the chemical potential $\mu$ for convenience~\cite{*[] [{, p29.}] lifshitz1980statistical} and choose $\mu$ to make the $N$-particle ground state have the lowest eigenenergy in the entire Hilbert space (Fock space).%~\footnote{For finite system at zero temperature $T=0$, the choice is not unique. One typical choice is to make $E_0^{N-1} - \mu(N-1) = E_0^{N+1} - \mu(N+1)$, i.e., $\mu = (E_0^{N+1} - E_0^{N-1})/2$, where $E_0^{N\pm 1}$ is the ground energy of the Hamiltonian $\hat{H}$ (without the chemical potential term) in the $N \pm 1$ particle sector. This choice is only equivalent to $\mu = 0$ at half-filling used in the Hubbard model Hamiltonian Eq.~(\ref{eq:hubbard}) (introduced in the next section) with even number of lattice sites. As a result, for odd number of sites, using $\mu = 0$ in Eq.~(\ref{eq:hubbard}) only guarantees the Fock space ground state being a half-filling state for $U > U_\text{cr}^L$. Numerically, we find $U_\text{cr}^L \approx 3.618 \tilde{t}$ for $L=3$ sites and $U_\text{cr}^L \approx 3.113 \tilde{t}$ for $L=5$ sites.}

In this work, we consider only a time-translationally invariant (time-homogeneous) system, so the Green's function only depends on $t-t'$ and we can choose $t' = 0$. In addition, if the Green's function is diagonal in spin space $G_{j\sigma, j'\sigma'}(t) = G_{j\sigma, j'\sigma}(t) \delta_{\sigma \sigma'}$ and spin-rotationally invariant $G_{j\ua, j'\ua}(t) = G_{j\da, j'\da}(t)$, we suppress the spin index when there is no confusion.

The double-time Green's function given in \cref{eq:GFt} can also be defined for more general $n$-body operators other than simple fermion creation and annihilation operators~\footnote{The generalization to higher order Green's functions, such as four-time (four-point) Green's functions, is out of the scope of this paper.}. This leads to the general retarded linear response function
\begin{align}
  G_{AB}^R (t) = -i\theta(t) \ev{ [\hat{A}(t), \hat{B}]_{\zeta} }{\psi_0^N},
  \label{eq:GABt}  
\end{align}
where $[\hat{A}(t), \hat{B}]_{\zeta} = e^{it\hat{H}'} \hat{A} e^{-it\hat{H}'}\hat{B} + \zeta \hat{B} e^{it\hat{H}'} \hat{A} e^{-it\hat{H}'}$, $\hat{H}' = \hat{H} - \mu \hat{N}$, and $\zeta=-1$ if $\hat{A}$ and $\hat{B}$ are bosonic and $\zeta=1$ if both are fermionic. The particle rank $n$ for the $n$-body operator $\hat{A}$ (or $\hat{B}$) is defined as the total number of (fermion or boson) creation and annihilation operators divided by two. If rank-$n$ operator $\hat{A}$ is a product of fermion creation and annihilation operators, $\hat{A}$ is bosonic for integer $n$ and fermionic for half-integer $n$.  \cref{eq:GABt} reduces to the single-particle Green's function \cref{eq:GFt} for rank-$\frac{1}{2}$ operators $\hat{A} = \hat{c}_{j\sigma}^\pdag$ and $\hat{B} = \hat{c}_{j'\sigma'}^\dagger$, while for the linear response to electromagnetic fields, such as charge and spin response functions, $\hat{A}$ and $\hat{B}$ are rank-$1$ operators given by a sum of products of two fermion (creation or annihilation) operators.

When the time domain Green's function $G_{AB}^R (t)$ is Fourier transformed to real-frequency domain, the resulted $G_{AB}^R (\omega)$ gives the dynamic response function. Methods for calculating dynamic response functions on quantum computers have been previously proposed, e.g., in Refs.~\onlinecite{Roggero19, Kosugi20linear}. In Ref.~\onlinecite{Roggero19}, the authors proposed calculating the linear response functions by constructing a perturbed state and then applying QPE to obtain the correct matrix elements for Green's functions. Their algorithm also gives access to the final states resulting from the perturbation, which could be useful for describing scattering experiments. In Ref.~\onlinecite{Kosugi20linear}, the authors similarly proposed using QPE with statistical sampling to measure the linear response functions, where the perturbed state is resulted from the action of a general electronic operator (i.e., $\hat{A}$ and $\hat{B}$ operators) implemented as a sum of unitary operators. In both works, the ground state before the perturbation is applied is either prepared by variational algorithms or given by an oracle. Drastically different from these works, our method for computing the Green's functions does not reply on QPE; instead, after preparing the ground state using our projective algorithm introduced above, we construct the resolvent operator with LCU to compute the Green's functions. The details and complexity analysis are given in the following section.

\subsection{Fourier-Laplace integral transform and {LCU} construction of resolvent operator}
The time domain Green's functions given in \cref{eq:GFt,eq:GABt} can be related to the real-frequency domain form by the Fourier-Laplace integral transform (FIT) with a convergence factor $e^{-\Gamma t}$ ($\Gamma > 0$) as follows.
\begin{align}
     G_{jj'}^R(\omega) 
  &= \int_{0}^{\infty} dt\,
     G_{jj'}^R(t) e^{i(\omega + i\Gamma)t} 
  \label{eq:GFFIT}
\\
  &= \ev**{\hat{c}_{j'}^\pdag
           R\qty(\omega_{+} + i\Gamma,
             \hat{H}')
           \hat{c}_{j}^\dagger}
          {\psi_0^N} +
     \ev**{\hat{c}_{j}^\dagger
           R\qty(\omega_{-} + i\Gamma,
             -\hat{H}')   
           \hat{c}_{j'}^\pdag}
          {\psi_0^N},
  \label{eq:GFRes}          
\end{align}
where the spin indices are suppressed, $\omega_{\pm} = \omega \pm \mu(\hat{N} - N)$, $\hat{H}' = \hat{H} - E_0^N$, $E_0^N$ is the $N$-particle ground energy, and the retarded resolvent operator is given by
\begin{align}
    R(\omega + i\Gamma, \hat{H}) 
  &= -i\int_{0}^{\infty} dt\,
     e^{i(\omega + i\Gamma-\hat{H})t}.
  \label{eq:ResFIT}
\end{align}
Performing the integral in \cref{eq:ResFIT} analytically we indeed obtain the usual definition of the resolvent operator,
\begin{align}
    R(\omega + i\Gamma, \hat{H}) 
  &= (\omega + i \Gamma - \hat{H})^{-1}.
  \label{eq:ResInv}
\end{align}
Discretizing the integral in \cref{eq:ResFIT} we obtain the following LCU approximation to the resolvent operator,
\begin{align}
     h(\omega + i\Gamma, \hat{H}) 
  &= -i\sum_{k=0}^{N_c} \Delta_t\,
     e^{i(\omega + i\Gamma - \hat{H})k\Delta_t}
   = \sum_{k=0}^{N_c} \alpha_k U_k,
  \label{eq:ResFITapprox}
\end{align}
where $\alpha_k = \Delta_t e^{-\Gamma k \Delta_t}$ and $U_k = e^{-i[(\hat{H} - \omega) k \Delta_t + \frac{\pi}{2}]}$. The computational complexity of constructing the resolvent operator \cref{eq:ResFIT} via the LCU approximant \cref{eq:ResFITapprox} is summarized in \cref{thm:res} and the proof is given in \cref{app:thm2}.

\begin{thm} \label{thm:res}
Consider a Hamiltonian $H$ with the spectrum $\sigma(\hat{H}) \subseteq [0, 1]$. Given $\abs{\omega} \in \sigma(\hat{H})$ and an artificial broadening $\Gamma$, the resolvent operator $R(\omega + i\Gamma, \hat{H})$ can be constructed via the LCU approximant \cref{eq:ResFITapprox} with an additive error within $\epsilon$, that is, $\norm*{R(\omega + i\Gamma, \hat{H}) -  h(\omega + i\Gamma, \hat{H})} \leq \epsilon$, if $N_c = \order{\frac{1}{\Gamma \epsilon} \log \frac{2}{\Gamma \epsilon}}$, $\Delta_t = \min\{\epsilon/2, 3/\norm*{\hat{H}}\} = \order{\epsilon}$. Implementing the LCU approximant \cref{eq:ResFITapprox} requires $\order*{\frac{1}{\Gamma^2} \log \frac{2}{\Gamma \epsilon}}$ queries to a time-evolution oracle for Hamiltonian $\hat{H}$ and $\log N_c$ ancilla qubits using the standard LCU formulation~\cite{LCU}.
\end{thm}

\subsection{Quantum circuit for resolvent operator and measurements of Green's function}
To simulate and measure the real-frequency domain Green's function $G_{jj'}^R(\omega)$ given in \cref{eq:GFRes} on quantum computer, we first apply the LCU approximant \cref{eq:ResFITapprox} to the resolvent,
\begin{subequations}
  \label{eq:ResAppr}
\begin{align}
  R(\omega_+ + i\Gamma, \hat{H}') &\approx 
  h_{+} \equiv \sum_{k=0}^{N_c} \alpha_k U_k^{+}
  = \sum_{k=0}^{N_c} 
    (\Delta_t e^{-\Gamma k \Delta_t})
    e^{-i[\hat{H} - E_0^N - \omega 
    -\mu(\hat{N} - N)] k \Delta_t
    -i\frac{\pi}{2}}, 
  \label{eq:ResAppra} \\
  R(\omega_- + i\Gamma, -\hat{H}') &\approx 
  h_{-} \equiv \sum_{k=0}^{N_c} \alpha_k U_k^{-}
  = \sum_{k=0}^{N_c} 
    (\Delta_t e^{-\Gamma k \Delta_t})
    e^{-i[-\hat{H} + E_0^N - \omega 
    +\mu(\hat{N} - N)] k \Delta_t
    -i\frac{\pi}{2}}.
  \label{eq:ResApprb} 
\end{align}
\end{subequations}

Next, we also need to express the fermion creation and annihilation operators, $\hat{c}_{j}^\dagger$ and $\hat{c}_{j}^\pdag$, in terms of unitary operators. This can be done through the following (Bogoliubov or Majorana) transformation. 
\begin{align}
    \hat{b}_{0j}^\pdag
  = \hat{c}_{j}^\pdag + \hat{c}_{j}^\dagger,
  \quad
    \hat{b}_{1j}^\pdag
  = i(\hat{c}_{j}^\pdag - \hat{c}_{j}^\dagger).
\end{align}
It is easy to verify that for $m\in \{0,1\}$, $\hat{b}_{mj}^\dagger = \hat{b}_{mj}^\pdag$ and $\hat{b}_{mj}^\pdag$ is unitary, i.e., $\hat{b}_{mj}^\dagger \hat{b}_{mj}^\pdag = 1$. Substituting $\hat{c}_{j}^\dagger = (\hat{b}_{0j}^\pdag + i\hat{b}_{1j}^\pdag)/2$, $\hat{c}_{j}^\pdag = (\hat{b}_{0j}^\pdag - i\hat{b}_{1j}^\pdag)/2$, and the LCU approximant for the resolvent $R(\omega_\pm + i\Gamma, \pm \hat{H}') \approx h_\pm$ into \cref{eq:GFRes}, we obtain
\begin{subequations}
  \label{eq:GF_LCU}
\begin{align}
    G_{jj'}^R(\omega)
  &\approx \ev**{
      \hat{c}_{j'}^\pdag h_{+}
      \hat{c}_{j}^\dagger +
      \hat{c}_{j}^\dagger h_{-}
      \hat{c}_{j'}^\pdag
     }{\psi_0^N}, \label{eq:GF_LCUa}\\
  &= \frac{1}{4} 
     \ev**{
     \begin{pmatrix}
     \phantom{+}
       \hat{b}_{0j'} h_{+} \hat{b}_{0j }
     +i\hat{b}_{0j'} h_{+} \hat{b}_{1j }
     -i\hat{b}_{1j'} h_{+} \hat{b}_{0j }
      +\hat{b}_{1j'} h_{+} \hat{b}_{1j }
      \\
      +\hat{b}_{0j } h_{-} \hat{b}_{0j'}
     -i\hat{b}_{0j } h_{-} \hat{b}_{1j'}
     +i\hat{b}_{1j } h_{-} \hat{b}_{0j'}
      +\hat{b}_{1j } h_{-} \hat{b}_{1j'}
     \end{pmatrix}
     }{\psi_0^N}. \label{eq:GF_LCUb}
\end{align}
\end{subequations}

\begin{figure} 
  \centering
  \setlength{\fboxsep}{0pt}
  \tikzsetnextfilename{fig_circuit}
  %\fbox{\import{texfigs/}{fig_circuit.tex}}
  \includegraphics[scale=1]{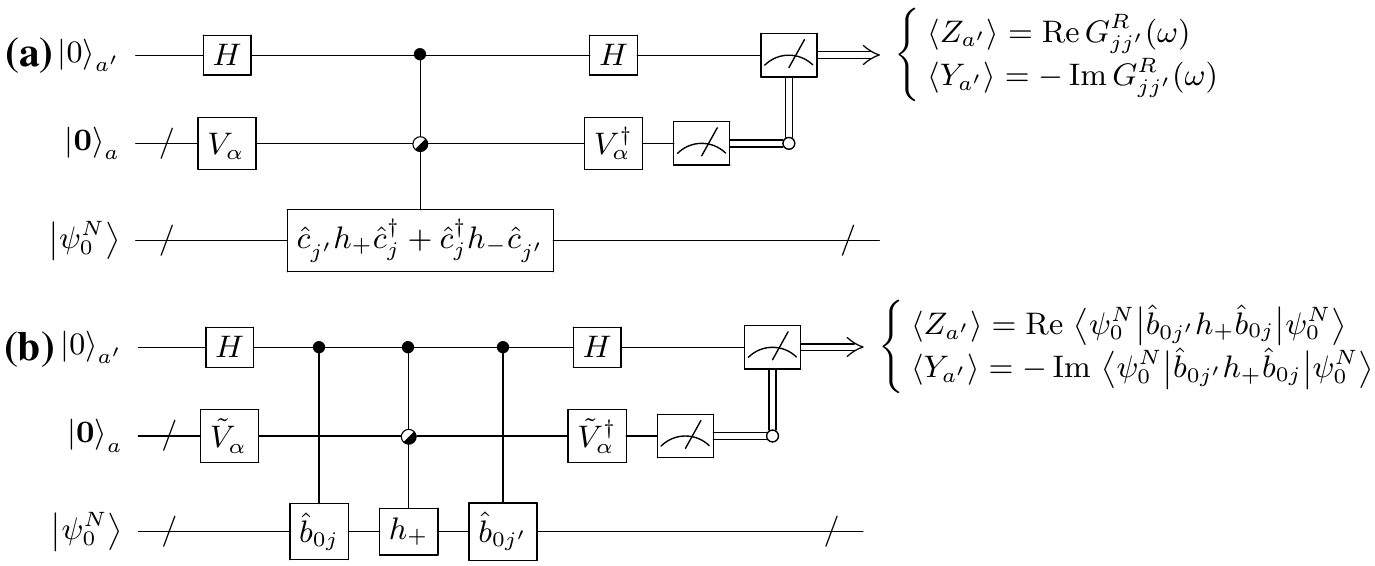}
\caption{(a) Circuit for the measurement of the frequency domain Green's function $G_{jj'}^R(\omega) = \ev{\qty\big (\hat{c}_{j'}^\pdag h_{+} \hat{c}_{j}^\dagger + \hat{c}_{j}^\dagger h_{-} \hat{c}_{j'}^\pdag)}{\psi_0^N}$ using the Hadamard test. (b) Circuit for the measurement of a subterm of $G_{jj'}^R(\omega)$, e.g., $\ev{\hat{b}_{0j'} h_{+} \hat{b}_{0j}}{\psi_0^N}$ using the Hadamard test. A single ancilla qubit $\ket{0}_{a'}$ is used for Hadamard test. The ancilla qubits $\ket{\mathbf{0}}_{a}$ are used for constructing LCU operators. $\tilde{V}_\alpha$ prepares the ancillary state with amplitudes given by the LCU coefficients in \cref{eq:ResAppra}, while $V_\alpha$ prepares a more complicated ancillary state to probabilistically implement the LCU operator $\qty\big (\hat{c}_{j'}^\pdag h_{+} \hat{c}_{j}^\dagger + \hat{c}_{j}^\dagger h_{-} \hat{c}_{j'}^\pdag)$. The half-black-half-white controls in (a) and (b) imply a network of controls and controls-on-zero on the ancilla qubits. The gate ($\hat{c}_{j'}^\pdag h_{+} \hat{c}_{j}^\dagger + \hat{c}_{j}^\dagger h_{-} \hat{c}_{j'}^\pdag$ or $h_{+}$) connected to this network of controls represents a sequence of controlled unitaries corresponding to individual terms of the LCU operators. The ancilla qubit for Hadamard test is only measured when the measured outcome of the LCU ancilla system state is $\ket{\mathbf{0}}_{a}$.}
  \label{fig:circuitGF}
\end{figure}

Since $h_{\pm}$ are LCU operators and all $\hat{b}_{mj}^\pdag$ are unitary operators, each of the eight individual operators $\hat{b}_{mj} h_{\pm} \hat{b}_{m'j'} = \sum_{k=0}^{N_c} \alpha_k (\hat{b}_{mj} U_k^{\pm} \hat{b}_{m'j'})$ inside the parentheses in \cref{eq:GF_LCUb} is also an LCU operator. Their sum gives an LCU expression for the operator $\qty\big(\hat{c}_{j'}^\pdag h_{+} \hat{c}_{j}^\dagger + \hat{c}_{j}^\dagger h_{-} \hat{c}_{j'}^\pdag)$ that completely determines the Green's function $G_{jj'}^R(\omega)$. This leads to two types of quantum circuits shown in \cref{fig:circuitGF}. The circuit in panel (a) implements the sum $\qty\big(\hat{c}_{j'}^\pdag h_{+} \hat{c}_{j}^\dagger + \hat{c}_{j}^\dagger h_{-} \hat{c}_{j'}^\pdag)$ as a single LCU operator, while the circuit in panel (b) implements individual LCU terms, for example, $\hat{b}_{0j'} h_{+} \hat{b}_{0j}$. The Green's function $G_{jj'}^R(\omega)$ is given by the expectation values of various LCU operators in the ground state. To measure such non-Hermitian operators, we use the straightforward Hadamard test as shown in \cref{fig:circuitGF}. In the Hadamard test, the operator to be measured is controlled on a single ancilla qubit and the real and imaginary expected values of the operator can be obtained by measuring the ancilla qubit.

As shown in \cref{fig:circuitGF}, to measure the Green's function, it involves implementing the controlled application of fermion operators $c_j^\pdag$ and $c_j^\dagger$ or rank-$n$ operators $\hat{A}$ and $\hat{B}$ in more general cases. When these operators are expressed in a linear combination of $\order{2^n}$ unitaries (each unitary can be a fermionic operator from the above Bogoliubov transformation or a Pauli string operator from the Jordan-Wigner transformation), the controlled application of the LCU operator can be implemented probablistically similar to \cref{fig:circuitGF}(a) and there is some quantum advantage in this case since $\order{n}$ ancilla qubits are sufficient even for high order (large $n$) correlation functions. However, the success probability of implementing the controlled application of $\hat{A}$ and $\hat{B}$ via LCU will decrease exponentially, and the best method to boost the success probability is still an open question. Since the $n$-body operator $\hat{A}$ (or $\hat{B}$) and its LCU representation are far from being unitary itself for high order correlation functions, methods such as oblivious amplitude amplification~\cite{Berry14} will fail in the extreme cases.

Alternatively, the controlled application of each individual term can be implemented deterministically similar to \cref{fig:circuitGF}(b) and the measured expectation values of all terms can be summed on classical computer. For individual terms represented by Pauli strings using the Jordan-Wigner encoding of fermion operators, the number of qubits acted on by the Pauli strings can be much larger than the rank of the operators $\hat{A}$ and $\hat{B}$. For specific hardware such as trapped ions, Ref.~\onlinecite{Pedernales2014} suggested using M{\o}lmer-S{\o}rensen gates to efficiently implement these controlled long Pauli strings.

\section{Applications and discussion}
In this section, we show numerical results using our algorithms for ground-state preparation and Green's function calculation for some of the important models in condensed matter physics, specifically the $q$-deformed XXZ chain in \cref{sec:FFqXXZ} and the one-dimensional Hubbard model in \cref{sec:hubbard}. The Hamiltonian for the $q$-deformed XXZ chain is frustration-free, while the Hamiltonian for the Hubbard model is frustrated. For the $q$-deformed XXZ chain, we only demonstrate the quadratic speedup of ground-state preparation when combining our projective method and the spectral gap amplification technique. For the Hubbad model, we give results on both ground-state preparation and Green's function calculation.

\subsection{$q$-deformed XXZ chain} \label{sec:FFqXXZ}
The frustration-free Hamiltonian for $q$-deformed XXZ chain~\cite{Wouters21, Gottstein1995, Alcaraz95} is 2-local and includes only nearest-neighbor spin-spin interactions. The Hamiltonian is given by $\hat{H} = \sum_{j=1}^{L-1} H_{j,j+1}$ for a $L$-site chain with an open boundary condition. The local term is given by~\cite{Gottstein1995}
\begin{align}
     H_{j,j+1} 
  &=  \frac{-q}{2(1 + q^2)} \qty(X_{j}X_{j+1} + Y_{j}Y_{j+1})
     +\frac{1}{4} \qty(1 - Z_{j}Z_{j+1})
     +\frac{1 - q^2}{4(1 + q^2)} \qty(Z_{j} - Z_{j+1}),
\end{align}
where $X_j$, $Y_j$, $Z_j$ are Pauli matrices and the parameter $q > 0$. The last terms in all $H_{j,j+1}$ ($1 \leq j \leq L-1$) add up to a boundary term $\frac{1 - q^2}{4(1 + q^2)} (Z_{1} - Z_{L})$ of $\hat{H}$. It is easy to verify $H_{j,j+1}^2 = 1$ so each $H_{j,j+1}$ is a projector .

For $q \neq 1$, the spectrum is gapped, including in the thermodynamic limit ($L \to \infty$)~\cite{Koma97}. The undeformed case $q=1$ is a fully isotropic ferromagnetic Heisenberg chain (the boundary term also vanishes) and it is well known that the spectral gap of isotropic ferromagnetic Heisenberg chain vanishes in the thermodynamic limit~\cite{Gottstein1995}. However, since the spectrum is discrete for a finite chain, it is gapped, i.e., the spectral gap $\Delta_s > 0$ for any $q > 0$. For simplicity, we choose a finite number $L \in \{4 ,6 ,8 ,10 \}$ and $q=1$ corresponding to the finite isotropic ferromagnetic Heisenberg chain, and use this model to demonstrate the quadratic speedup of ground-state preparation with our projective algorithm combined with the spectral gap amplification~\cite{Chowdhury2017, Somma2013}. Since each local term $H_{j,j+1}$ is a projector, it is straightforward to apply the spectral gap amplification algorithm given in Ref.~\onlinecite{Chowdhury2017}. Specifically, the spectral-gap amplified Hamiltonian $\hat{H}_r$ introduced in \cref{sec:GapAmp} has the following block matrix form in the computational basis $\hat{H}_r = \spmqty{\bm{0} &\bm{\Pi} \\ \bm{\Pi}^\dagger &\bm{0}}$, where $\bm{\Pi} = \pmqty{H_{1,2} &H_{2,3} &\dots &H_{L-1,L}}$ is a row-vector of sub-block matrices. We use this matrix representation in numerical calculations, but in quantum simulations, we can use the following equivalent form $\hat{H}_r = \sum_{j=1}^{L-1} \qty( \op{0}{j}_{a_r} + \op{j}{0}_{a_r}) \otimes H_{j,j+1}$, which satisfies \cref{eq:HSGapAmp} relating $\hat{H}$ and its ``square-root'' $\hat{H}_r$. Therefore, the minimal number of ancilla qubits to construct $\hat{H}_r$ is $\ceil{\log_2 L}$.

Another reason we consider the isotropic case is that for $q=1$ the $(L+1)$-fold degenerate ground states are the $L+1$ Dicke states~\cite{Zhou2011} $\ket{D_j^L} = \sqrt{j!(L-j)!/L!} \sum_{|x| = j} \ket{x}$, where $0 \leq j \leq L$, $x \in \{0, 1\}^L$ is a bit string, and $|x|$ is the Hamming weight defined as the number of ones in $x$. The sum $\sum_{|x| = j}$ includes all bit strings with the same Hamming weight $j$. Dicke states have applications in areas such as quantum metrology and quantum computing, and various probabilistic and deterministic methods have been proposed to prepare Dicke states in quantum systems~\cite{Zhou2011, Wang2021} or on quantum computers~\cite{Bartschi2019, Mukherjee2020}. Since LCU is a core routine in our algorithm, our projective state preparation method is probabilistic. To prepare a Dicke state $\ket{D_j^L}$ with Hamming weight $j$, we choose a simple initial state $\ket{x_0} = \ket{1 \cdots 1 0 \cdots 0}$ with the Hamming weight $|x_0| = j$. For such an initial state, we will consider preparing the state $\ket{D_j^L}$ with $j = L/2$ below. This is the most challenging case since the overlap between the initial state and the true ground state is smaller than any other different $j$.

\begin{figure} 
  \centering
  \includegraphics[width=0.9\columnwidth, keepaspectratio]{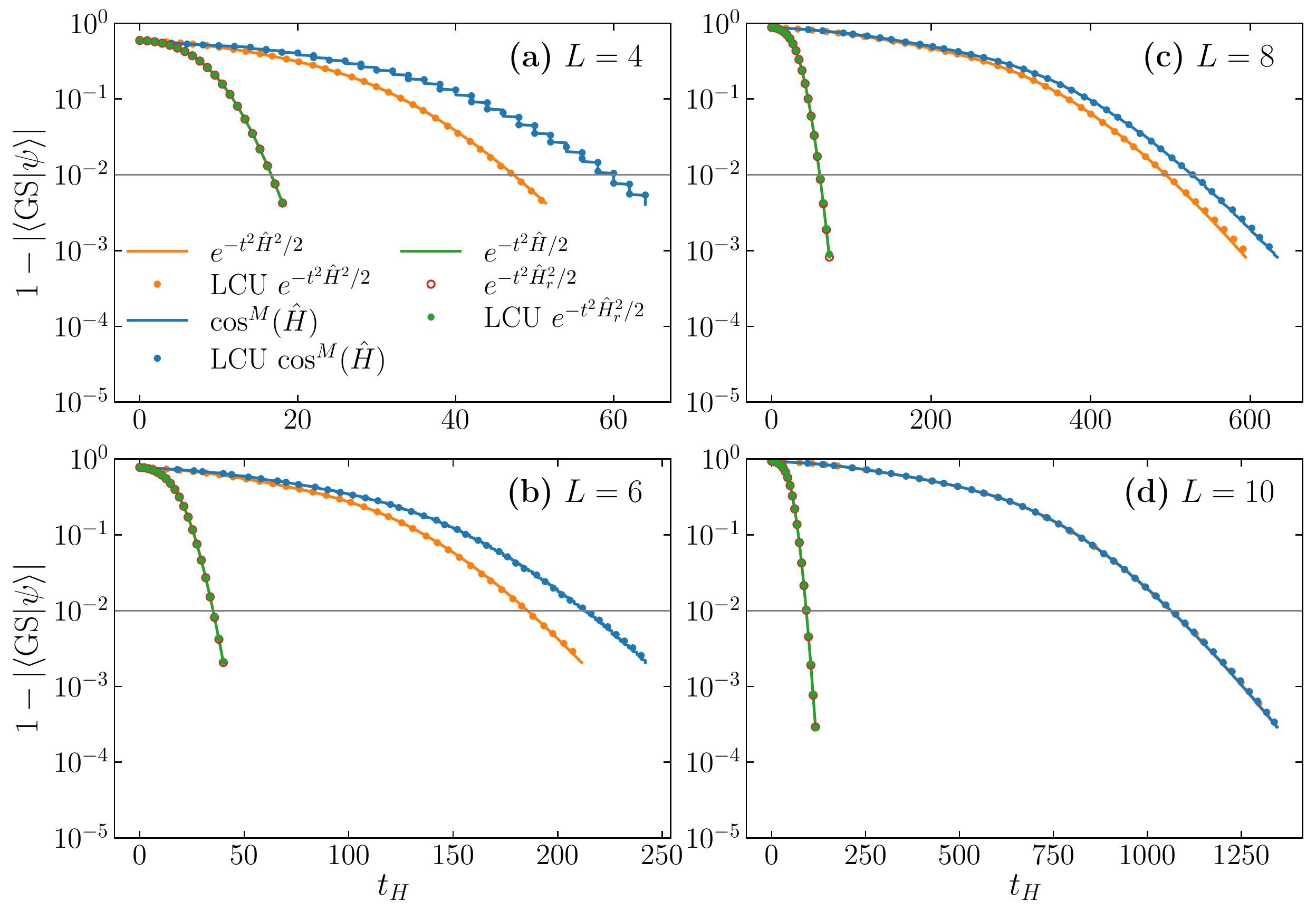}
\caption{Projective ground-state preparation for isotropic ferromagnetic Heisenberg chain with (a) $L=4$ , (b) $L=6$, (c) $L=8$, and (d) $L=10$ sites. The true ground state is the Dicke state $\ket{D^L_{L/2}}$ and the initial state used is $\ket{x_0} = \ket{1 \dots 1 0 \dots 0}$ with Hamming weight $|x_0| = L/2$. Each panel compares the query complexity of three different projective ground-state preparation algorithms: (i) $e^{-\frac{1}{2} t^2 \hat{H}^2}$ operator, (ii) $\cos^{M}(\hat{H})$ operator, (iii) $e^{-\frac{1}{2} t^2 \hat{H}_r^2}$ operator with spectral gap amplified Hamiltonian $\hat{H}_r$. The Hamiltonian is renormalized and its spectrum lies between 0 and 1. The results from LCU (dots) implementation using \cref{thm:lcuHST} are plotted alongside the results from the exact implementation using sparse matrix representation (solid lines and red circles for $e^{-\frac{1}{2} t^2 \hat{H}_r^2}$). $t_H$ quantifies the query complexity to Hamiltonian time-evolution oracle and is defined as the longest effective time evolved by the oracle. The full definition of $t_H$ is given in the main text. The gray horizontal line marks the targeted fidelity error $\epsilon = 0.01$ for which we set various parameters given in \cref{thm:lcuHST}.}
  \label{fig:fids_qXXZ}
\end{figure}

In \cref{fig:fids_qXXZ}, we compare the query complexity of three different projective ground-state preparation algorithms: (i) our algorithm using $e^{-\frac{1}{2} t^2 \hat{H}^2}$ operator, (ii) algorithm of Ge \textit{et al.}~\cite{Ge2019} using $\cos^{M}(\hat{H})$ operator, (iii) our algorithm with spectral gap amplification using $e^{-\frac{1}{2} t^2 \hat{H}_r^2}$. The results from LCU implementation of these operators given by \cref{thm:lcuHST} are plotted alongside the results from the exact implementation using sparse matrix representation. We quantify the query complexity with the parameter $t_H$ defined as the longest effective time evolved by the Hamiltonian time-evolution oracle. For our algorithm (i) and (iii), $t_H = t z_c = t \sqrt{2 \log \frac{2}{\gamma \eta}}$, where $t$ is varied from 0 to the the lower bound $\frac{1}{\Delta} \sqrt{2\log \frac{1}{\gamma\eta}}$ given by \cref{thm:lcuHST}, which is the lower bound required to ensure the fidelity error within $\epsilon = \eta^2/2$ (we used $\epsilon = 0.01$). In (i) we used $\Delta = \Delta_s$, which is the true spectral gap of $\hat{H}$, and in (iii) we used $\Delta = \sqrt{\Delta_s}$, which is the spectral gap of $\hat{H}_r$. For algorithm (ii) of Ge \textit{et al.}, $t_H = 2\times \min \qty{\ceil*{\sqrt{\frac{M}{2} \log \frac{2}{\gamma\eta}}}, \frac{M}{2}}$, where the even integer $M$ is varied from 0 to the lower bound $2 \ceil*{\frac{1}{\Delta^2} \log \frac{1}{\gamma \eta}}$ that is required to ensure the fidelity error within $\epsilon$. Note that we used the tightened bounds derived in \cref{app:ge_bounds} in the algorithm (ii) of Ge \textit{et al.}; taking the integer value in the $t_H$ definition causes the zig-zag appearance in some of results from algorithm (ii) of Ge \textit{et al.}, for example, \cref{fig:fids_qXXZ}(a), the blue line and the blue dots.

From \cref{fig:fids_qXXZ}, we see algorithm (iii), our method combined with the spectral gap amplification, gives the best scaling. For larger system size with decreasing normalized spectral gap, the query complexity reduction can be quite significant. As a side note, our algorithm (i) and the algorithm (ii) of Ge \textit{et al.} show similar scalings and the difference becomes smaller as the system size grows.

\subsection{Hubbard model} \label{sec:hubbard}
The Hamiltonian of the Hubbard model is given by 
\begin{align}
     \hat{H}
  &= -\tilde{t} \sum_{\la i,j \ra, \sigma}
     \qty(\hat{c}_{i\sigma}^\dagger \hat{c}_{j\sigma}^\pdag
         +\hat{c}_{j\sigma}^\dagger \hat{c}_{i\sigma}^\pdag)
     +U\sum_i \qty(\hat{n}_{i\ua} - \frac{1}{2})
              \qty(\hat{n}_{i\da} - \frac{1}{2})
     -\mu\sum_{i,\sigma}\hat{n}_{i\sigma},
  \label{eq:hubbard}
\end{align}
where $\la i,j \ra$ denotes nearest-neighbor sites $i$ and $j$, $\hat{c}_{i\sigma}^\dagger$ ($\hat{c}_{i\sigma}^\pdag$) creates (annihilates) an electron on site $i$ with spin $\sigma \in \{\ua, \da \}$, $\hat{n}_{i\sigma} = \hat{c}_{i\sigma}^\dagger \hat{c}_{i\sigma}^\pdag$ is the electron number density operator, $\tilde{t}$ is the nearest-neighbor hopping integral, $\mu$ is the chemical potential, and $U$ is the local Hubbard repulsion between electrons.

Although it is one of the most easily expressible models in condensed matter physics that includes interactions between electrons, the Hubbard model captures many interesting physical phenomena, such as the Mott metal-insulator transition~\cite{Mott1968, Imada1998, Georges1992}, antiferromagnetism~\cite{White1989}, emergent spin and stripe orders~\cite{Zheng1155, Huang2018}, strange metallic behavior~\cite{Huang987}, pseudogaps~\cite{Gull2013, Chen2015}, and high-temperature superconductivity~\cite{Maier2005, Gull2013}, depending on its dimensionality and parameter regime. Despite its simple form, the Hubbard model only has exact solutions in one or infinite dimensions. Therefore, most of the progress in understanding the Hubbard model is gained through advanced numerical simulations or approximate many-body theoretical techniques. However, simulations carried out on classical computers suffer from some sort of exponential scaling in, e.g., memory requirements or computation time, due to the exponential growth of Hilbert space and, more fundamentally, the notorious negative sign problem~\cite{Troyer2005}. Quantum computers promise to alleviate the exponential scaling of the memory requirement by storing the many-body quantum states in a number of qubits that only grows linearly with the system size.

In this work, we apply our algorithms to prepare the ground state and compute the single-particle Green's function of the one-dimensional Hubbard model with $L$ lattice sites and the periodic boundary condition. We will compare the results for $L \in \{2, 3, 4, 5\}$ for which the number of fermion modes is $2L$ (factor of $2$ from two spin species) and the dimensions of the Hilbert space are $2^{2L}$.  We also choose the chemical potential so that the average particle number density $n = \frac{1}{L}\sum_{i\sigma} \ev{\hat{n}_{i\sigma}} = 1$, i.e., the system is at half-filling. We have written the Hamiltonian \cref{eq:hubbard} in a particle-hole symmetric form, as indicated by the subtraction of $\frac{1}{2}$ in the interaction term proportional to $U$. For this form, the chemical potential $\mu = 0$ at half-filling. We consider a strong correlation case with an electron repulsion strength of $U/\tilde{t} = 8$.

\subsubsection{Ground-state preparation}
The choice of initial trial state is crucial to the success of projective state preparation procedure. For the strong interaction $U$ we choose, a single-component antiferromagnetic product state is a good initial trial state. This corresponds to the fermionic state $\ket{\ua_{0}, \da_{1}, \cdots, \ua_{L}, \da_{L-1}}$ for even $L$ (the state $\ket{\ua_{0}, \da_{1}, \cdots, \ua_{L-1}}$ for odd $L$), where each lattice site is singly-occupied by either up- or down-spin electrons in a staggered pattern. At $U/\tilde{t} = 8$ the interaction strength imposes the large penalty for doubly occupied configurations which our input state avoids.
    
\begin{figure} 
  \centering
  \includegraphics[width=0.9\columnwidth, keepaspectratio]{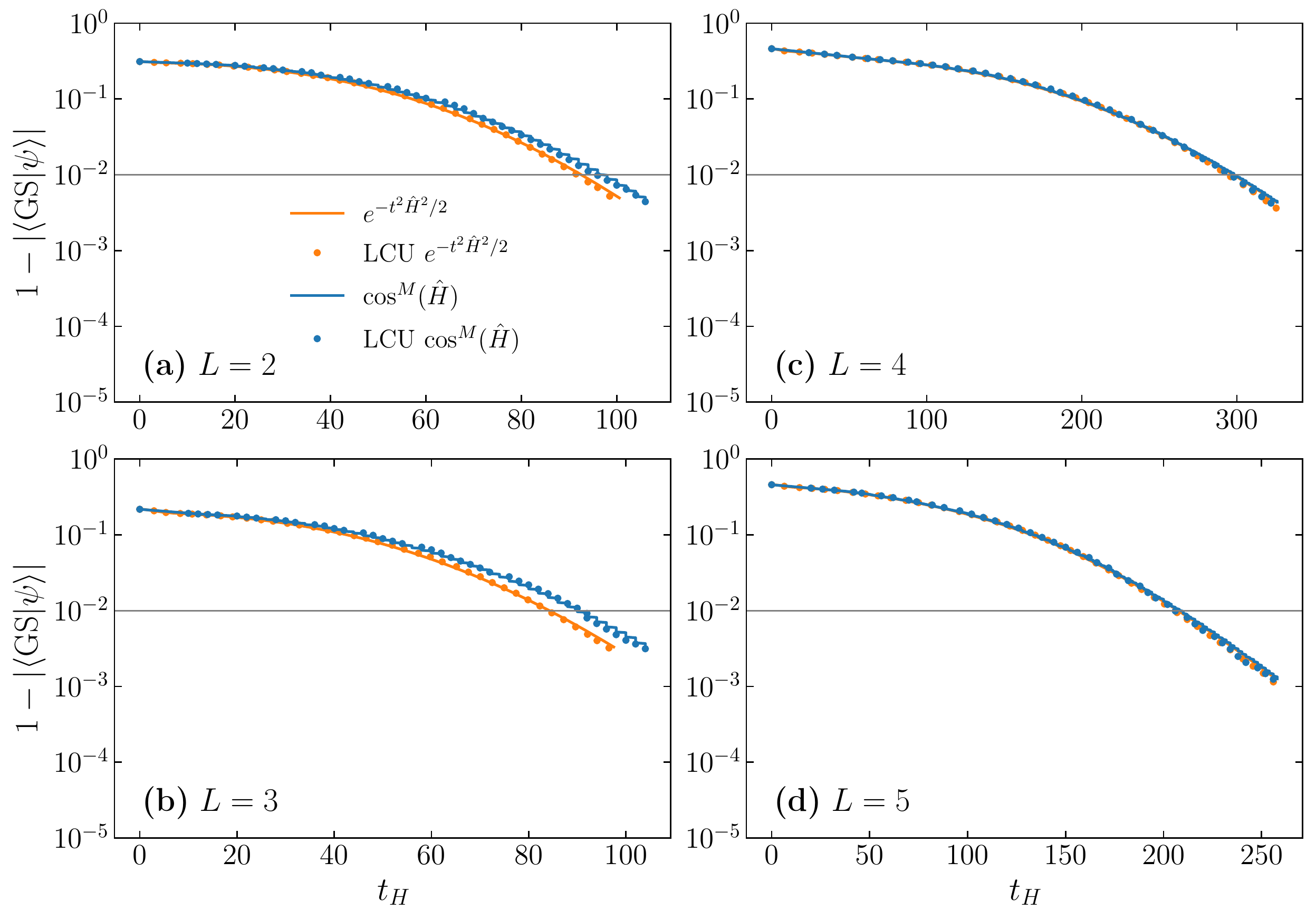}
\caption{Projective ground-state preparation for Hubbard model \cref{eq:hubbard} with (a) $L=2$ , (b) $L=3$, (c) $L=4$, and (d) $L=5$ sites. The parameters $\tilde{t} = 1$, $U=8$, and at half-filling $\mu = 0$. Each panel compares the query complexity of two different projective ground-state preparation algorithms: (i) $e^{-\frac{1}{2} t^2 \hat{H}^2}$ operator and (ii) $\cos^{M}(\hat{H})$ operator. The Hamiltonian has been renormalized and its spectrum lies between 0 and 1. The results from LCU (dots) implementation using \cref{thm:lcuHST} are plotted alongside the results from the exact implementation using sparse matrix representation (solid lines). $t_H$ is defined in the same way as in \cref{fig:fids_qXXZ}. The gray horizontal line marks the targeted fidelity error $\epsilon = 0.01$ for which we set various parameters given in \cref{thm:lcuHST}.}
  \label{fig:fids}
\end{figure}

In \cref{fig:fids}, we compare the query complexity of two different projective ground-state preparation algorithms: (i) our algorithm using $e^{-\frac{1}{2} t^2 \hat{H}^2}$ operator, and (ii) algorithm of Ge \textit{et al.}~\cite{Ge2019} using $\cos^{M}(\hat{H})$ operator and the tightened bounds derived in \cref{app:ge_bounds}. For each algorithm, the parameter $t_H$ quantifying the query complexity is defined above in \cref{sec:FFqXXZ}. Similar to the XXZ chain, we find our algorithm performs slightly better than the algorithm of Ge \textit{et al.}~\cite{Ge2019} in smaller system sizes. By the panel of \cref{fig:fids}(d) both algorithms show almost identical scalings, which agrees with the asymptotic query complexity given in \cref{tab:scaling} for these two algorithms.

\begin{figure}
  \centering
  \includegraphics[width=0.9\columnwidth, keepaspectratio]{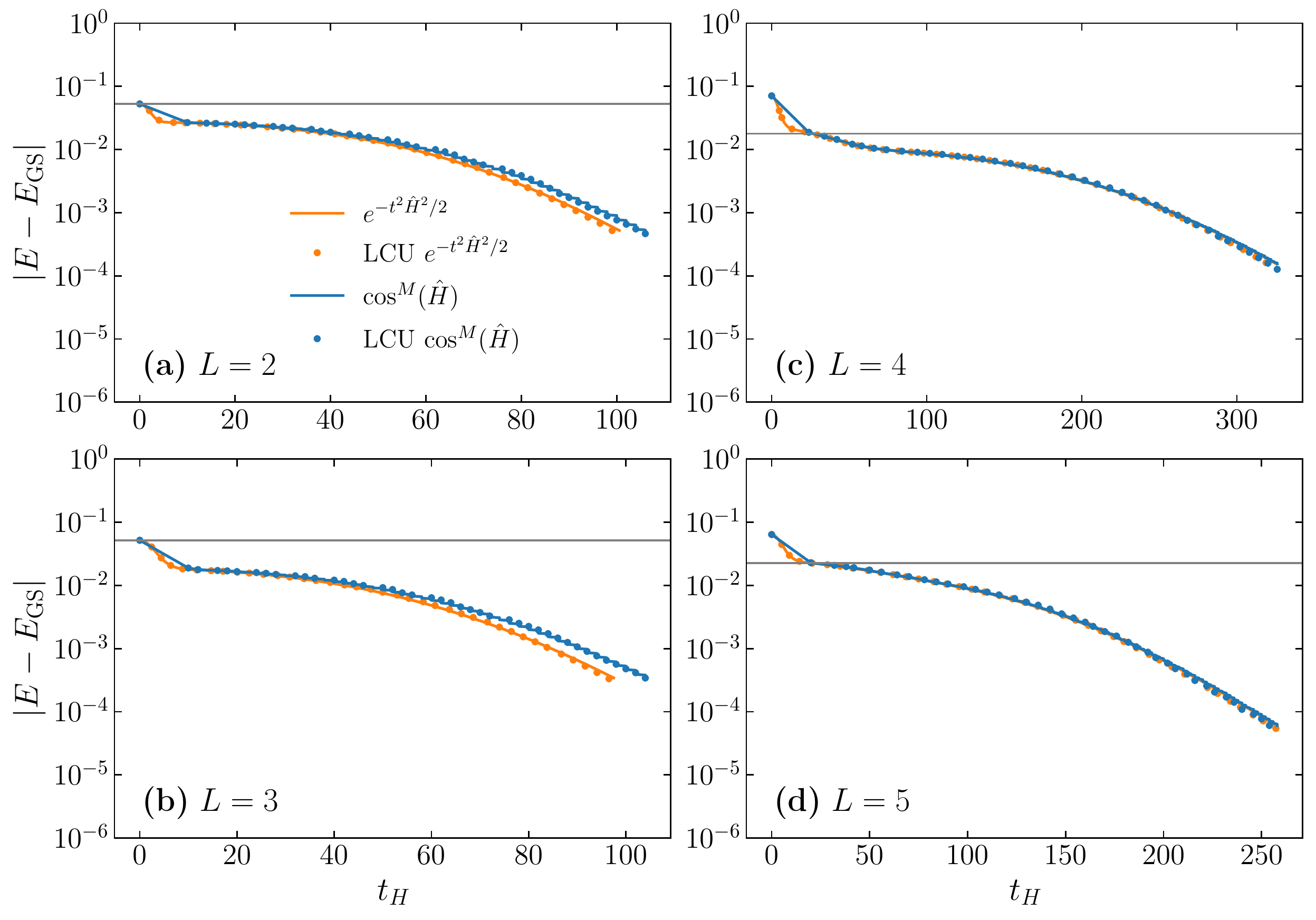}
\caption{Additive error of the ground state energy of the Hubbard model \cref{eq:hubbard} with (a) $L=2$ , (b) $L=3$, (c) $L=4$, and (d) $L=5$ sites. The ground state used in calculating the energy expectation values are those shown in \cref{fig:fids}. Each panel compares the query complexity of two different projective ground-state preparation algorithms: (i) $e^{-\frac{1}{2} t^2 \hat{H}^2}$ operator and (ii) $\cos^{M}(\hat{H})$ operator. The horizontal gray line denotes the value of the true spectral gap $\Delta_s$.}
  \label{fig:GSE}
\end{figure}

In \cref{fig:GSE}, we plot the difference $\abs{E - E_\text{GS}}$ between the approximate ground energy $E$ using the prepared ground state and the true ground energy $E_\text{GS}$. Note that the Hamiltonian is renormalized and its spectrum lies between 0 and 1, so the true ground energy $E_\text{GS}$ is set to exactly zero (within machine precision in our numerical simulation and, for quantum simulation, within the required precision given in \cref{tab:scaling}). We notice that the additive error $\abs{E - E_\text{GS}}$ for the ground energy in \cref{fig:GSE} computed via both our algorithm and the algorithm of Ge \textit{et al.}, approaches the chosen $\epsilon=0.01$ more quickly than the error in the fidelity does in \cref{fig:fids}. This is likely due to the fact that our choice of trial initial state is an excellent trial state with respect to energy for the strong interaction $U$ we chose.

%%%%%%%%%%

\subsubsection{Single-particle Green's function calculation}
\begin{figure} 
  \centering
  \includegraphics[width=0.9\columnwidth, keepaspectratio]{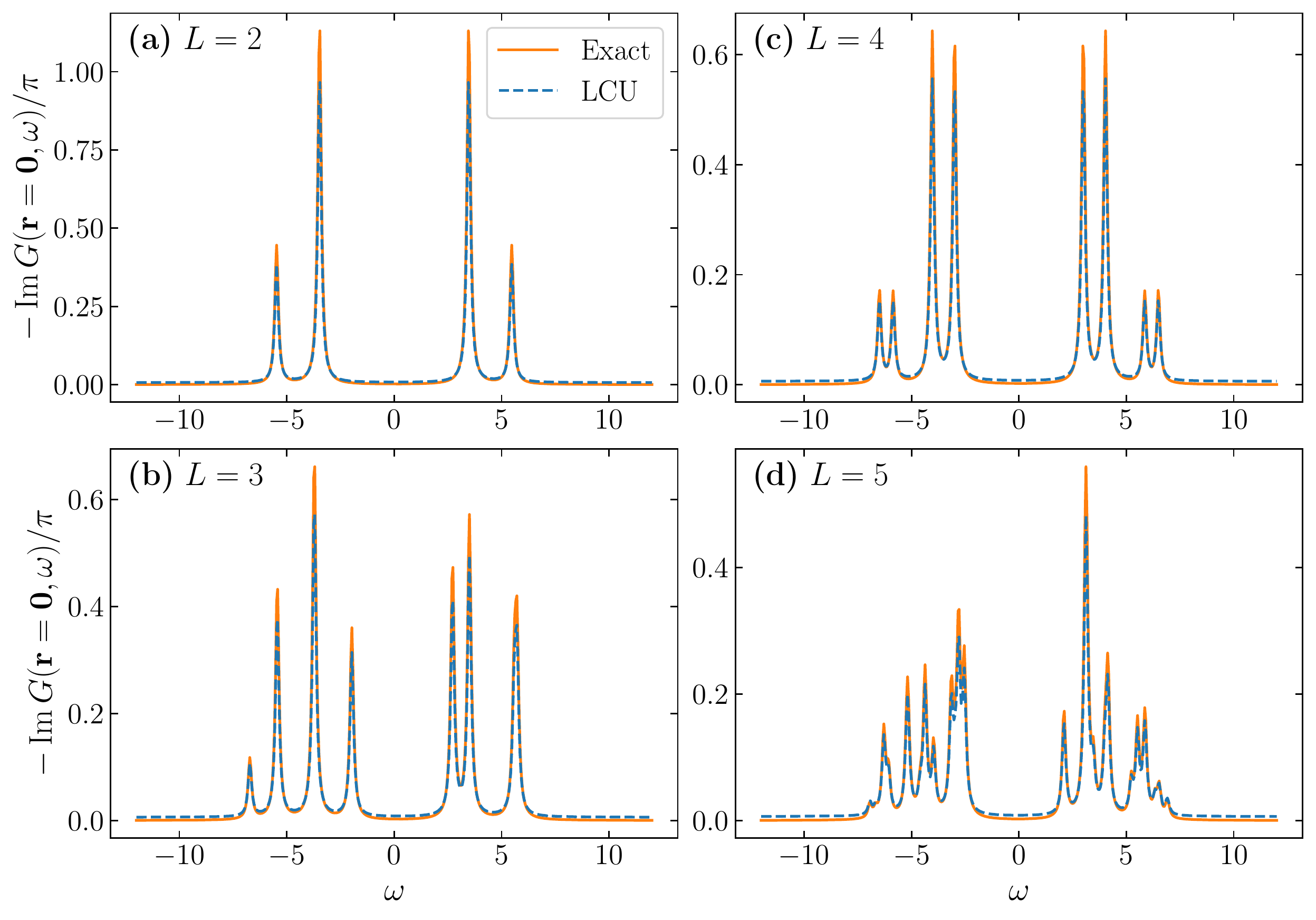}
\caption{Calculations of the local density of states for the first lattice site of the two, three, four, and five site Hubbard model with parameters $t = 1$ and $U=8$ at half-filling. Here, we have constructed the resolvent operator both exactly (orange) and via LCU (blue), traced over the degenerate ground states of the 3 and 5-site models, and renormalized each to 1. LCU refers to the linear combination of unitaries approximant to the Laplace transform construction of the resolvent operator given in \cref{eq:ResFITapprox}.}
  \label{fig:g00}
\end{figure}

In the Hubbard model, the single particle Green's function in the frequency domain can be rewritten from \cref{eq:GFRes} as 
\begin{align} \label{eq:GFRes2}
     G_{jj'}^R(\omega) 
  &= \sum_{\alpha}
     \frac{(M_{j'}^{\alpha})^* M_{j}^{\alpha}}
          {\omega + \mu - E_\alpha^{N+1} + E_0^N + i\delta} +
     \sum_{\beta}
     \frac{(L_{j}^{\beta})^* L_{j'}^{\beta}}
          {\omega + \mu + E_\beta^{N-1} - E_0^N + i\delta},
\end{align}
where $E_\alpha^{N \pm 1}$ is the eigenenergy of the $\psi_\alpha^{N \pm 1}$ eigenstate of the $N \pm 1$ particle sector, $\mu$ is the chemical potential which we set to 0, $E_0^N$ is the ground state energy of the $N$ particle sector, and $\delta$ is a convergence factor from the Fourier integral transform of the time domain Green's function.
The weights of the poles in the Green's function are given by
\begin{align}
     M_{j}^{\alpha} 
  &= \mel*{\psi^{N+1}_{\alpha}}
          {\hat{c}_{j}^\dagger}
          {\psi_0^N} 
   = \mel*{\psi_0^N}
          {\hat{c}_{j}^\pdag}
          {\psi^{N+1}_{\alpha}}^*,
  \quad
     L_{j}^{\beta} 
   = \mel*{\psi^{N-1}_{\beta}}
          {\hat{c}_{j}^\pdag}
          {\psi_0^N}
   = \mel*{\psi_0^N}
          {\hat{c}_{j}^\dagger}
          {\psi^{N-1}_{\beta}}^*.
\end{align}
In our numerical simulations, we utilize \cref{eq:GFRes2} to compute the Green's function and local density of states. However, on the quantum computer the Green's function will be calculated with equation \cref{eq:GF_LCUa} (probabilistically) or \cref{eq:GF_LCUb} (deterministically). This discrepancy is due to convenience for the classical numerical simulations and the fact that, 
for our numerical simulations, it suffices to show that the resolvent operator can be constructed via the discretized FIT given in \cref{eq:ResFITapprox}. To show this, we calculate the resolvent via \cref{eq:ResFITapprox} and use it to compute the local Green's function of the first lattice site, i.e. $G_{00}^R(\omega)$, using \cref{eq:GFRes2}. In \cref{fig:g00}, we plot the local density of states (LDOS) for Hubbard chains of size $2$--$5$ corresponding to panels (a)--(d). We see that for a conservative allowable error of $\epsilon' = 0.05$ and broadening $0.1$, we are able to reproduce the relevant peaks and their relative weights in the local density of states. For the degeneracy in the ground state for the 3 and 5-site cases, we trace over the degenerate ground states, and we renormalize the local density of states for all cases. Our choice of strong electron interactions ($U/\tilde{t} = 8$) allows us to clearly see the gap in the local density of states characteristic of a Mott insulating phase.

\section{Conclusion} \label{sec:conclusion}
We have presented quantum algorithms for ground state preparation and response function calculation. Our projective method for preparing the ground state of a system based on the Hubbard-Stratonovich transformation of an imaginary time like operator. Our state preparation algorithm matches the optimal scaling in the spectral gap $\Delta$ and initial overlap $\gamma$ between trial and ground states~\cite{Lin2020}. However, our algorithm quadratically reduced the precision to which a ground state estimate must be known (compared to that reported in Ref.~\citenum{Ge2019}), but matches the scaling of Ref.~\citenum{Ge2019} when their bounds are tightened as shown in Appendix \ref{app:ge_bounds}. 

Our algorithm for computing the response function of a system via construction of the resolvent through the discretized Fourier-Laplace transform using LCU shows good agreement with the exact value and tractable scaling. Our algorithm is the first proposed to directly construct the resolvent operator on a quantum computer. It does not rely on variational principles or statistical sampling, but instead directly constructs the resolvent operator integral transformations and LCU.

To verify our algorithms and determine the algorithm's complexity in practice we have performed numerical simulations for both ground state preparation and computation of the Green's function in the context of the paradigmatic Fermi-Hubbard model in one dimension with different numbers of sites. We have also performed numerical simulations for ground state preparation of the q-deformed XXZ model, since the Fermi-Hubbard model does not satisfy the frustration free requirement of spectral gap amplification. We find that our method for ground state preparation and that of Ref.~\citenum{Ge2019} rapidly asymptotically converge even for relatively small system sizes. 

% future work and open questions 

\begin{acknowledgments}
The authors thank Ryan Bennink, Jim Freericks, Alexander Kemper, and Oleksandr Kyriienko for valuable comments and discussions. T.K. is supported by the U.S.~Department of Energy, Office of Science, Office of Workforce Development for Teachers and Scientists, Office of Science Graduate Student Research (SCGSR) program. The SCGSR program is administered by the Oak Ridge Institute for Science and Education (ORISE) for the DOE. ORISE is managed by ORAU under contract number DE-SC0014664. %
E.F.D. acknowledges DOE ASCR funding under the Quantum Computing Application Teams program, FWP number ERKJ347. %
Y.W. acknowledges DOE ASCR funding under the Quantum Application Teams program, FWP number ERKJ335.
\end{acknowledgments}

\appendix

\section{Proofs} \label{sec:proofs}

\subsection{Proof of \cref{lem:tbound}} \label{app:tboundProof}
Consider a Hamiltonian $\hat{H}$ with a spectral representation $\hat{H} = \sum_{l} \lambda_l \op{\lambda_l}{\lambda_l}$, where all eigenvalues $\lambda_l \geq 0$. Denote the ground state $\ket{\lambda_0}$ and the first excited state $\ket{\lambda_1}$, and the spectral gap $\Delta_s = \lambda_1 - \lambda_0 > 0$. Given a normalized initial state $\ket*{\psi_0}$ that has overlap with ground state $|\ip{\lambda_0}{\psi_0}|=\gamma>0$, we derive a lower bound for the parameter $t$ in the operator $f(\hat{H}) = e^{-\frac{1}{2} t^2\hat{H}^2}$ so that the projected state $\ket*{\tilde{\psi}} = f(\hat{H}) \ket*{\psi_0}$, after the normalization $\ket*{\psi} = \ket*{\tilde{\psi}}/\norm*{\ket*{\tilde{\psi}}}$, is $\epsilon$-close to the ground state. That is, the infidelity $1 - |\ip{\lambda_0}{\psi}| < \epsilon$, and, equivalently, the Euclidean distance $\norm*{\ket{\psi} -\ket{\lambda_0}} <\sqrt{2\epsilon} \equiv \eta$.

Expand $\ket*{\psi_0}$ in the eigenbasis: $\ket*{\psi_0} = c_0 \ket{\lambda_0} + \sum_{l>0} c_l \ket{\lambda_l}$. For degenerate ground states, $\hat{H} \ket{\lambda_0^\alpha} = \lambda_0 \ket*{\lambda_0^\alpha}$, the initial state is given by $\ket*{\psi_0} = \sum_{\alpha} c_0^\alpha \ket*{\lambda_0^\alpha} + \sum_{l>0} c_l \ket{\lambda_l} = c_0 \ket{\lambda_0} + \sum_{l>0} c_l \ket{\lambda_l}$, where $c_0 \equiv (\sum_\alpha |c_0^\alpha|^2)^{1/2}$ and $\ket{\lambda_0} \equiv \sum_{\alpha} (c_0^\alpha/c_0) \ket*{\lambda_0^\alpha}$.

Note that although we define $\Delta_s$ to be the exact spectral gap $\lambda_1 - \lambda_0$ and $\gamma$ the exact overlap $|c_0|$ between the initial state and the ground state, lower bounds of these parameters can be used when the exact values are not known \textit{a priori} and all complexity bounds proved here still hold, just not as tight as when exact values are used.

\begin{align}
    \ket*{\tilde{\psi}}
  &= f(\hat{H}) \ket*{\psi_0}
   = e^{-\frac{1}{2} t^2\hat{H}^2}\ket{\psi_0}
  \label{eq:fvectnm1}\\
  &= c_0 e^{-\frac{1}{2} t^2\lambda_0^2}  
     \qty[\ket{\lambda_0} + \sum_{l > 0} \frac{c_l}{c_0}
     e^{-\frac{1}{2} t^2(\lambda_l^2 - \lambda_0^2)}
     \ket{\lambda_l} ],
  \label{eq:fvectnm2}\\
     \ket*{\psi}
  &= \ket*{\tilde{\psi}}/\norm*{\ket*{\tilde{\psi}}}. \\
  %%%%%
     1 - |\ip{\lambda_0}{\psi}|
  &= 1 - \left[
        1 + \sum_{l > 0} \frac{|c_l|^2}{|c_0|^2} 
        e^{-t^2(\lambda_l^2 - \lambda_0^2)}
        \right]^{-1/2}
  \label{eq:fvectnm3} \\
  &\leq 1 - \left[
        1 + e^{-t^2(\lambda_1^2 - \lambda_0^2)}\sum_{l > 0} \frac{|c_l|^2}{|c_0|^2} 
        \right]^{-1/2} \\
  &= 1 - \left[
        1 + e^{-t^2(\lambda_1^2 - \lambda_0^2)}\frac{1 - \gamma^2}{\gamma^2} 
        \right]^{-1/2} \\
  &< 1 - \left[
        1 - \frac{1}{2}e^{-t^2(\lambda_1^2 - \lambda_0^2)}\frac{1 - \gamma^2}{\gamma^2} 
        \right] \\
  &= e^{-t^2(\lambda_1^2 - \lambda_0^2)}\frac{1 - \gamma^2}{2\gamma^2} \label{eq:fin_lem1}.
\end{align}

To make the infidelity $1 - |\ip{\lambda_0}{\psi}| < \epsilon$, it is sufficient to set
\begin{align}
  & e^{-t^2(\lambda_1^2 - \lambda_0^2)}\frac{1 - \gamma^2}{2\gamma^2}
  = e^{-t^2[(\lambda_0 + \Delta_s)^2 - \lambda_0^2]}\frac{1 - \gamma^2}{2\gamma^2}
  \leq e^{-t^2[(\lambda_0 + \Delta)^2 - \lambda_0^2]}\frac{1 - \gamma^2}{2\gamma^2}
  < \epsilon \\
  \implies 
t &> \frac{1}{\Delta} 
     \qty[\log ( \frac{1-\gamma^2}{2\gamma^2 \epsilon})]^{1/2}
     \qty(1 + \frac{2\lambda_0}{\Delta})^{-1/2} \\
  &= \frac{1}{\Delta}
     \qty[2\log(\frac{1}{\gamma \eta }) + \log(1-\gamma^2)]^{1/2}
     \qty(1 + \frac{2\lambda_0}{\Delta})^{-1/2} \\
  &= \order{\frac{1}{\Delta} \sqrt{2 \log \frac{1}{\gamma \eta}}}.
\end{align}

The final expression has been given in a form to easily compare with Ge \textit{et al.}~\cite{Ge2019} involving the typical factors $\frac{1}{\Delta}$ and $\log\frac{1}{\gamma \eta}$. 

\subsection{Discretization and truncation}
\label{app:d&t}
Now we proceed to give the conditions to satisfy $\norm*{h(\hat{H}) - f(\hat{H})} \leq \epsilon' = \gamma \eta$. Since $\norm*{h(\hat{H}) - f(\hat{H})} \leq \norm*{h(\hat{H}) - h^{\infty}(\hat{H})} + \norm*{h^{\infty}(\hat{H}) - f(\hat{H})} \equiv \epsilon_t + \epsilon_d$, where the first term is the truncation error and the second discretization error. Here, $h^{\infty}(\hat{H})$ is the infinite sum without truncation at $k=\pm N_z$. Now we find the conditions so that $\epsilon_t +\epsilon_d < \epsilon'$; and we find the optimal conditions in the balanced case $\epsilon_t = \epsilon_d < \epsilon'/2$.

The truncation error $\epsilon_t < e^{-z_c^2/2}$ can be shown as follows.
\begin{align}
     \epsilon_t 
  &= \norm{h(\hat{H}) - h^{\infty}(\hat{H})} \\
  &= \norm{\frac{1}{\sqrt{2\pi}} \sum_{|k|=N_z + 1}^{\infty}
     \Delta_z e^{-\frac{1}{2}k^2 \Delta_z^2} e^{-i t k\Delta_z\hat{H}}} \\
  &\leq \frac{2}{\sqrt{2\pi}} \sum_{k=N_z + 1}^{\infty} 
     \Delta_z e^{-\frac{1}{2}k^2 \Delta_z^2} \\
  &< \frac{2}{\sqrt{2\pi}} \int_{N_z \Delta_z}^{\infty}dz\,
     e^{-\frac{1}{2} z^2}
   = \frac{2}{\sqrt{\pi}} \int_{z_c/\sqrt{2}}^{\infty}dx\,
     e^{-x^2} 
     \quad (z_c\equiv N_z \Delta_z)\\
  &\leq \frac{e^{-z_c^2/2}}
    {\frac{\sqrt{\pi}z_c}{2\sqrt{2}} +
     \sqrt{\qty(\frac{\sqrt{\pi}z_c}{2\sqrt{2}})^2 + 1}} \\
  &= \exp(-\frac{z_c^2}{2} - \arcsinh\frac{\sqrt{\pi}z_c}{2\sqrt{2}})
   = \exp(-\frac{z_c^2}{2} 
     - \frac{1}{2}\arccosh \qty(\frac{\pi z_c^2}{4} + 1) ) \\
  &\equiv e^{-z'^2_c/2} < e^{-z^2_c/2},
\end{align}
where $z_c' \equiv \qty [z_c^{2} + \arccosh (\pi z_c^2/4 + 1)]^{1/2}$. In the above, we used the fact that $\sum_{k=N_z + 1}^{\infty}$ is the right Riemann sum of the integral $\int_{N_z \Delta_z}^{\infty}dz$ of a monotonically decreasing function and hence is an underestimation. In the third to the last step, the inequality used to bound the integral is from Ref.~\onlinecite{*[] [{, Eq.~(7.1.13).}] Milton_Abramowitz1964}.

$\forall b\in (-a, a)$, we define $M$ as an upper bound of the following integral
\begin{align}
     \int_{-\infty}^{\infty} dx\, \norm{f(x + ib)} 
  &= \frac{1}{\sqrt{2\pi}}\int_{-\infty}^{\infty} dx\,
     \norm{e^{-\frac{1}{2}(x+ib)^2}e^{-it(x+ib)\hat{H}}}  \\
  &= \frac{1}{\sqrt{2\pi}}\int_{-\infty}^{\infty} dx\,
     e^{-\frac{1}{2}x^2} e^{\frac{1}{2}b^2} \norm{e^{tb\hat{H}}} \\
  &= e^{\frac{1}{2}b^2} \norm{e^{tb\hat{H}}}
   \leq  e^{\frac{1}{2}b^2 + t|b|}
   \leq  e^{\frac{1}{2}a^2 + ta} \equiv M.
\end{align}
We have assumed the spectrum of Hamiltonian operator $\sigma(\hat{H})\in [0, 1]$ so the spectrum norm $\norm*{\hat{H}} \leq 1$. Then, by the Theorem 5.1 of Ref.~\onlinecite{trefethen14}, the discretization error is
\begin{align}
     \epsilon_d 
  &= \norm{h^{\infty}(\hat{H}) - f(\hat{H})} \\
  &\leq \frac{2M}{e^{2\pi a/\Delta_z} - 1}
   = \frac{2e^{\frac{1}{2}a^2 + ta}}{e^{2\pi a/\Delta_z} - 1} \\
  & \approx 2\exp\left(\frac{1}{2}a^2 + ta - \frac{2\pi a}{\Delta_z}\right) \\
  &= 2\exp\left[-a\left(\frac{2\pi}{\Delta_z}- t - \frac{a}{2}\right)\right].
\end{align}

Now, we solve for $a$ that minimizes the above for $2\pi/\Delta_z > t$. That is, we find $a$ that maximizes $a \left(\frac{2\pi}{\Delta_z} - t - \frac{a}{2} \right)$ for $a>0$. So $a = \frac{2\pi}{\Delta_z} - t$ and $\epsilon_d \leq e^{-[(2\pi/\Delta_z) -t]^2/2}$. At this optimal condition, the balance condition $\epsilon_d = \epsilon_t$ leads to $(2\pi/\Delta_z) - t = z'_c$, i.e., the step size $\Delta_z = 2\pi / (z'_c + t)$. Therefore, $N_z = z_c / \Delta_z =  z_c(z'_c + t)/2\pi$. Finally,  $\epsilon_t = \epsilon_d < \epsilon'/2 = \gamma \eta/2$ gives $z_c = \order{ \sqrt{\log\frac{1}{\gamma \eta}} }$.

\subsection{Proof of Theorem \ref{thm:lcuHST}}
\label{app:thm1}
So far we have proved that $1 - |\ip{\lambda_0}{\psi}| < \epsilon$, provided $t >\frac{1}{\Delta_s} \sqrt{2 \log \frac{1}{\gamma \eta}}$. Combining this bound with \cref{eq:fvectnm1,eq:fvectnm2,eq:fvectnm3}, we have
\begin{align}
  &\gamma e^{-\frac{1}{2} t^2\lambda_0^2}
  \leq
  \norm{f(\hat{H}) \ket*{\psi_0}}
  \leq
  \gamma e^{-\frac{1}{2} t^2\lambda_0^2} \frac{1}{1 - \epsilon}
   \\
  \implies
  & \norm{f(\hat{H}) \ket*{\psi_0}}
  = \gamma e^{-\frac{1}{2} t^2\lambda_0^2} (1 + \order*{\eta^2}).
\end{align}

If we use the discretized and truncated LCU formula $h(\hat{H})$ instead of $f(\hat{H})$, we will prove below that given $t = \order{\frac{1}{\Delta_s}\sqrt{\log \frac{1}{\gamma \eta }} }$, $|\lambda_0| < \delta_0 = \order{\Delta_s / \sqrt{\log\frac{1}{\gamma\eta}}}$, and $\norm*{h(\hat{H}) - f(\hat{H})} < \epsilon' = \gamma \eta$ then the state prepared by applying $h(\hat{H})$ is $\epsilon$-close to the ground state, 
\begin{align} \label{bnd1}
  \norm{\frac{h(\hat{H})\ket{\psi_0}}
        {\norm*{h(\hat{H})\ket{\psi_0}}}
        - \ket{\lambda_0}} 
  &\leq \eta (1 + 2e^{\frac{1}{2} t^2\lambda_0^2})
   = \order{\eta}.
\end{align}
Here, $\delta_0$ is equivalent to the precision of the ground energy $\tilde{\lambda}_0$ known \textit{a priori} for a given Hamiltonian $\tilde{H}$. Before executing our algorithm, we shift $\tilde{H}$ to $\hat{H}$ by a constant so that the same ground state now has the ground energy $\lambda_0$ satisfying $|\lambda_0| < \delta_0$. First, for any two vectors $\ket{u}$ and $\ket{v}$,
\begin{align}
     \norm{\frac{\ket{u}}{\norm{\ket{u}}}
          -\frac{\ket{v}}{\norm{\ket{v}}}}
  &= \norm{\frac{\ket{u} - \ket{v}}{\norm{\ket{u}}}
          +\frac{\ket{v}}{\norm{\ket{u}}}
          -\frac{\ket{v}}{\norm{\ket{v}}}} \\
  &\leq \frac{\norm{\ket{u} - \ket{v}}}{\norm{\ket{u}}}
       +\frac{\norm{\ket{v}(\norm{\ket{v}}-\norm{\ket{u}})}}
             {\norm{\ket{u}}\norm{\ket{v}}} \\
  &= \frac{\norm{\ket{u} - \ket{v}}}{\norm{\ket{u}}}
    +\frac{\abs{\norm{\ket{u}}-\norm{\ket{v}}}}
          {\norm{\ket{u}}} \\
  &\leq \frac{\norm{\ket{u} - \ket{v}}}{\norm{\ket{u}}}
       +\frac{\norm{\ket{u} - \ket{v}}}{\norm{\ket{u}}} \\
  &= \frac{2\norm{\ket{u} - \ket{v}}}{\norm{\ket{u}}}.\label{eq:uvdiff}
\end{align}
In the second to the last step, we used the reverse triangle inequality. Similarly, we have $\norm{\ket{u}/\norm{\ket{u}} - \ket{v}/\norm{\ket{v}}} \leq 2\norm{\ket{u} - \ket{v}}/\norm{\ket{v}}$. Then,
\begin{align}
  \norm{\frac{h(\hat{H})\ket{\psi_0}}
        {\norm*{h(\hat{H})\ket{\psi_0}}}
        - \ket{\lambda_0}} 
  &= \norm{
       \frac{h(\hat{H})\ket{\psi_0}}
            {\norm*{h(\hat{H})\ket{\psi_0}}}
      -\frac{f(\hat{H})\ket{\psi_0}}
            {\norm*{f(\hat{H})\ket{\psi_0}}}
      +\frac{f(\hat{H})\ket{\psi_0}}
            {\norm*{f(\hat{H})\ket{\psi_0}}}        
      -\ket{\lambda_0}} \\
  &\leq 
   \norm{
       \frac{h(\hat{H})\ket{\psi_0}}
            {\norm*{h(\hat{H})\ket{\psi_0}}}
      -\frac{f(\hat{H})\ket{\psi_0}}
            {\norm*{f(\hat{H})\ket{\psi_0}}}} +
   \norm{
       \frac{f(\hat{H})\ket{\psi_0}}
            {\norm*{f(\hat{H})\ket{\psi_0}}}
      -\ket{\lambda_0}} \\
  &\leq \frac{2\norm*{[h(\hat{H})-f(\hat{H})]\ket{\psi_0}}}
             {\norm*{f(\hat{H})\ket{\psi_0}}}
        +\eta \label{eq:diff1} \\
  &\leq \frac{2\gamma\eta}{\gamma e^{-\frac{1}{2}t^2 \lambda_0^2}}
       +\eta \\
  &= \label{eq:diff2}\eta (1 + 2e^{\frac{1}{2} t^2\lambda_0^2})
   = \order{\eta}.
\end{align}
In the second to the last step, we used $\norm*{[h(\hat{H}) - f(\hat{H})] \ket{\psi_0}} \leq \norm*{h(\hat{H}) - f(\hat{H})} < \gamma \eta$ and $\norm*{f(\hat{H})\ket{\psi_0}} \geq \gamma e^{-\frac{1}{2}t^2 \lambda_0^2}$. In the last step, we used $t \lambda_0 = \order{1}$.

Denote the normalized ground state prepared by the LCU operator as $\ket{\psi} = h(\hat{H})\ket{\psi_0} / \norm*{h(\hat{H})\ket{\psi_0}}$, the error bound on the fidelity can be derived as follows.
\begin{align}
    1 - \abs{\ip{\lambda_0}{\psi}} 
  &\leq \abs{1 - \ip{\lambda_0}{\psi}} \\
  &= \abs{\ip{\lambda_0} - \ip{\lambda_0}{\psi})}
   = \abs{\bra{\lambda_0}(\ket{\lambda_0} - \ket{\psi})} \\
  &\leq \norm{\ket{\lambda_0}} \norm{\ket{\lambda_0} - \ket{\psi}} \\
  &= \norm{\ket{\lambda_0} -
  \frac{h(\hat{H})\ket{\psi_0}}{\norm*{h(\hat{H})\ket{\psi_0}}}} \\
  &\leq \eta ( 1 + 2e^{\frac{1}{2}t^2 \lambda_0^2} ).
\end{align}
In the first step above, we used the reverse triangle inequality $1 - \abs{z} \leq \abs{1 - z}$, $\forall z \in \mathbb{C}$. The second inequality used is  Cauchy-Schwarz inequality $\abs{\ip{u}{v}} \leq \norm{u} \norm{v}$. The last inequality is \cref{bnd1}. Considering the general bounds for fidelity $0\leq 1 - \abs{\ip{\lambda_0}{\psi}} \leq 1$ for any two normalized vectors, the above upper bound $\eta ( 1 + 2e^{\frac{1}{2}t^2 \lambda_0^2} )$ is only useful if $\eta ( 1 + 2e^{\frac{1}{2}t^2 \lambda_0^2} ) < 1$. In this case, a tighter error bound $(\eta^2/2) ( 1 + 2e^{\frac{1}{2}t^2 \lambda_0^2} )^2$ is derived as follows.
\begin{align}
    1 - \abs{\ip{\lambda_0}{\psi}} 
  &\leq \frac{1}{2} (2 - 2\Re \ip{\lambda_0}{\psi}) \\
  &= \frac{1}{2} (\ip{\lambda_0} + \ip{\psi}
     - \ip{\lambda_0}{\psi} - \ip{\psi}{\lambda_0}) \\
  &= \frac{1}{2}\norm{\ket{\psi} -\ket{\lambda_0}}^2 \\
  &\leq \frac{\eta^2}{2}( 1 + 2e^{\frac{1}{2}t^2 \lambda_0^2} )^2.
\end{align}

Finally, the query complexity is $\alpha t z_c / \gamma = \order{\frac{\alpha}{\gamma\Delta_s} \log\frac{1}{\gamma \eta} }$, where the factor of $1/\gamma$ comes from the minimum label finding algorithm (or amplitude amplification) in Ref.~\citenum{Ge2019} to get an estimate for the ground state energy. This is the same complexity as Ge \textit{et al.}~\cite{Ge2019} and Lin and Tong~\cite{Lin2020} for ground state preparation (ground energy known within a bound $\delta$). But our bound $\delta = \order{ \Delta_s / \sqrt{\log\frac{1}{\gamma\eta}} }$, derived in Appendix \ref{app:ge_bounds},  quadratically reduces the precision requirement reported in Ge \textit{et al.}~\cite{Ge2019} of $\delta = \order{ \Delta_s / \log\frac{1}{\gamma\eta} }$.

A short summary of parameters to set in numerical calculations.
\begin{itemize}
  \item $\gamma = |\ip{\lambda_0}{\psi_0}|$
  \item $\epsilon = 1 - |\ip{\lambda_0}{\psi_t}|$;
        $\eta = \norm*{\ket{\psi_t} - \ket{\lambda_0}}  \approx \sqrt{2\epsilon}$
  \item $t = \order{ \frac{1}{\Delta_s} \sqrt{\log\frac{1}{\gamma \eta}}}$
  \item $z_c = \order{  \sqrt{\log\frac{1}{\gamma \eta}}}$
  \item $\Delta_z = 2\pi / (z'_c + t)$; $N_z = [ z_c /  \Delta_z]$
\end{itemize}

\subsection{Proof of Theorem \ref{thm:res}}
\label{app:thm2}
First, let us confirm that through the Fourier-Laplace integral transform (FIT) we do in fact recover the resolvent operator. 
\begin{align}
     R(\omega+i\Gamma,\hat{H})
  &= -i \int_{0}^{\infty} \dd{t} e^{i(\omega+ i\Gamma)t}e^{-it\hat{H}} 
   = -i \int_{0}^{\infty} \dd{t} e^{i(\omega - \hat{H} + i\Gamma)t}
  \quad \text{($\Gamma > 0$)}
\\
  &= -i \left.
    \frac{e^{i(\omega - \hat{H})t} e^{-\Gamma t}}{i(\omega - \hat{H} + i\Gamma)}
    \right|_{0}^{\infty}
   = -i
     \frac{e^{i(\omega - \hat{H})\infty} e^{-\Gamma \infty} -
           e^{i(\omega - \hat{H})0} e^{-\Gamma 0}}
          {i(\omega - \hat{H} + i\Gamma)}
\\
  &= \frac{1}{\omega - \hat{H} + i\Gamma}. 
\\[2em] %%%%%%%%%%%%%%%%%%%%
\end{align}
Next, we discretize the FIT to express it as an LCU.
\begin{align}
     h(\omega+i\Gamma,\hat{H})
  &= -i \sum_{k=0}^{N_c} \Delta_t\,
     e^{i(\omega - \hat{H} + i\Gamma)k\Delta_t} \\
  &= -i \sum_{k=0}^{N_c} \Delta_t 
     \qty[e^{i(\omega - \hat{H} + i\Gamma)\Delta_t}]^{k} \\
  &= -i \Delta_t 
     \frac{1 - e^{i(\omega - \hat{H} + i\Gamma)\Delta_t(N_c + 1)}}
          {1 - e^{i(\omega - \hat{H} + i\Gamma)\Delta_t}} \\
  \mycomment{
  &= \Delta_t  
     \frac{1 - e^{i(\omega - \hat{H} + i\Gamma)\Delta_t(N_c + 1)}}
     {e^{i(\omega - \hat{H} + i\Gamma)\Delta_t/2}
      \qty[e^{-i(\omega - \hat{H} + i\Gamma)\Delta_t/2} - e^{i(\omega - \hat{H} + i\Gamma)\Delta_t/2}]}
  }
  &\approx \frac{i \Delta_t}{ e^{i(\omega - \hat{H} + i\Gamma)\Delta_t} - 1}
  \quad \text{($e^{-\Gamma\Delta_t (N_c + 1)} = 
              e^{-\Gamma t_c} e^{-\Gamma\Delta_t}\to 0$
              as $t_c=N_c\Delta_t \to \infty$)} \\
  &\approx \frac{i \Delta_t}{i(\omega - \hat{H} + i\Gamma)\Delta_t}
  \quad \text{($e^{x}-1\to x$ as
              $x =i(\omega - \hat{H} + i\Gamma)\Delta_t \to 0$ as
              $\Delta_t \to 0$)} \\
  &= \frac{1}{\omega - \hat{H} + i\Gamma} = R(\omega+i\Gamma,\hat{H}).
\\ %%%%%%%%%%%%%%%%%%%%
\end{align}
We then borrow the general method and notation of Ref.~\citenum{trefethen14} and bound the truncation and discretization error.
\begin{align}
    \norm{h(\omega+i\Gamma,\hat{H}) - R(\omega+i\Gamma,\hat{H})} 
  & \leq \norm{h(\omega+i\Gamma,\hat{H}) - h^{\infty}(\omega+i\Gamma,\hat{H})} +
         \norm{h^{\infty}(\omega+i\Gamma,\hat{H}) - R(\omega+i\Gamma,\hat{H})}
  \equiv \epsilon_t + \epsilon_d.
\end{align}

Suppose we want to limit the sum of truncation and discretization errors to  $\epsilon_t + \epsilon_d = \epsilon'$. We can choose the balanced case $\epsilon_t = \epsilon_d = \frac{\epsilon'}{2}$. For truncation error,
\begin{align}
     \epsilon_t 
  &= \norm{h(\omega+i\Gamma,\hat{H}) - h^{\infty}(\omega+i\Gamma,\hat{H})}
   = \norm{\sum_{k=N_c + 1}^{\infty}\Delta_t
     e^{i(\omega - \hat{H} + i\Gamma)k\Delta t}} \\
  &\leq \sum_{k=N_c + 1}^{\infty}\Delta_t
     \norm{e^{i(\omega - \hat{H})k\Delta t}}
     e^{-\Gamma k\Delta t}
  \quad \text{($e^{-\Gamma k\Delta t} < 1$
              $\forall {\Gamma > 0, \Delta_t > 0, k>0}$)} \\
    %I added in this step for my own clarity.
    &= \Delta_t \frac{\norm{e^{i (\omega - H)(N_c + 1) \Delta_t}}e^{-\Gamma(N_c+1)\Delta_t}}{1-\norm{e^{i(\omega - H)\Delta_t}}e^{-\Gamma \Delta_t}} \\
  &= \frac{\Delta_t e^{-\Gamma (N_c + 1)\Delta_t}}
          {1 - e^{-\Gamma \Delta_t}}
   = \frac{\Delta_t e^{-\Gamma N_c \Delta_t}}
          {e^{\Gamma \Delta_t} - 1} 
  \quad \text{($N_c \Delta_t = t_c$)}
\\
  &= \frac{\Gamma\Delta_t}{e^{\Gamma \Delta_t} - 1}
     \frac{e^{-\Gamma t_c}}{\Gamma}
   < \frac{e^{-\Gamma t_c}}{\Gamma}.
\end{align}
In the last step, we have used $\frac{x}{e^x - 1} = x/(x + \frac{1}{2!}x^2 + \dots) < 1$, $\forall x=\Gamma\Delta_t > 0$. Solving $\epsilon_t < \frac{e^{- \Gamma t_c}}{\Gamma} = \frac{\epsilon'}{2}$, we find
\begin{align}
     t_c 
  &= \frac{1}{\Gamma} \log\qty(\frac{2}{\Gamma \epsilon'}), \\
     N_c
  &= \frac{t_c}{\Delta_t}
   = \frac{1}{\Gamma \Delta_t} \log\qty(\frac{2}{\Gamma \epsilon'}).
\end{align}

Now we inspect the discretization error $\epsilon_d$.
\begin{align}
     \epsilon_d 
  &= \norm{h^{\infty}(\omega+i\Gamma,\hat{H}) - R(\omega+i\Gamma,\hat{H})} \\
  &= \norm{
     \frac{i \Delta_t}{e^{i(\omega - \hat{H} + i\Gamma)\Delta_t} - 1} -
     \frac{1}{\omega - \hat{H} + i\Gamma}} \\
  &= \Delta_t \norm{\frac{1}{e^{z} - 1} - \frac{1}{z}}
  \quad (z = i(\omega - \hat{H} + i\Gamma)\Delta_t) \\
  &= \Delta_t \norm{-\frac{1}{2} + \frac{z}{12} + \order{z^3}} \\
  &< \frac{1}{2}\Delta_t
    +\frac{2W}{12}\Delta_t^2
    +\norm*{\order{\Delta_t^4}}. \\
     W
  &\equiv \norm*{\omega - \hat{H} + i\Gamma}/2 \\
  &\lesssim \abs*{\lambda_\text{max}(\hat{H}) - \lambda_\text{min}(\hat{H})}/2
  \quad (\forall \omega \in \qty[\lambda_\text{min}(\hat{H}), \lambda_\text{max}(\hat{H})], \Gamma \ll 
  \max\qty{\abs*{\lambda_\text{min}(\hat{H})}, \abs*{\lambda_\text{max}(\hat{H})}})
  \notag \\
  &\leq (\abs*{\lambda_\text{max}(\hat{H})} +
         \abs*{\lambda_\text{min}(\hat{H})})/2
  \notag\\
  &< (\norm\big{\hat{H}}_s + \norm\big{\hat{H}}_s)/2 = \norm\big{\hat{H}}_{s},
\end{align}
where the single-particle excitation spectrum norm is defined as (use the original form of $\hat{H}\to \hat{H}-E_0^N$ in $G^{>}$ resolvent):
\begin{align}
    \norm\big{\hat{H}}_{s} &\equiv 
    \sup_{\alpha} \norm\big{\hat{H}\ket{\psi^{N + 1}_{\alpha}}}
  = \sup_{\alpha} \abs\big{E_{\alpha}^{N + 1} - E_0^N}.  
\end{align}

Now we solve for $\Delta_t$ as follows.
\begin{align}
  & \epsilon_d
  < \frac{1}{2}\Delta_t +\frac{2W}{12}\Delta_t^2 +\norm*{\order{\Delta_t^4}} 
  = \frac{\epsilon'}{2} \\
  \Longrightarrow
  & \Delta_t + \frac{W}{3}\Delta_t^2 +\norm*{\order{\Delta_t^4}} = \epsilon'. \\
\intertext{Let $\Delta_t = \epsilon' + a \epsilon'^2 + b \epsilon'^3 + \order{\epsilon'^4}$,}
  & \epsilon' + a \epsilon'^2 + b \epsilon'^3 +
    \frac{W}{3}\epsilon'^2 (1 + 2a\epsilon') + \order{\epsilon'^4} = \epsilon' \\
  & \epsilon' + (a + \frac{W}{3})\epsilon'^2 + 
    (b + \frac{2aW}{3})\epsilon'^3 + \order{\epsilon'^4} = \epsilon' \\
  \Longrightarrow
  &a = -\frac{W}{3},\ b = \frac{2W^2}{9}. \\
  &\Delta_t = \epsilon' - \frac{W}{3}\epsilon'^2 + \frac{2W^2}{9}\epsilon'^3.
\end{align}
The bound on $\Delta_t$ is not useful in practice since the parameter $W \equiv \norm*{\omega - \hat{H} + i\Gamma}$ is not easy to compute and it depends on $\omega$. However, since $W\leq \norm*{H}$, we verify that the choice $\Delta_t = \min\{\epsilon'/2, 3/\norm*{H}\}$ suffices to ensure $\epsilon_d < \epsilon'/2$ if $\norm*{\order{\Delta_t^4}}$ can be neglected. Note that $\Delta_t \leq \epsilon'/2$ and $\Delta_t \leq 3/\norm*{H}$; the latter gives $W\Delta_t \leq 3$.
\begin{align}
  \epsilon_d
  &< \frac{1}{2}\Delta_t +\frac{2W}{12}\Delta_t^2
    +\norm*{\order{\Delta_t^4}} \\
  &\approx \frac{\Delta_t}{2}\qty(1 + \frac{W \Delta_t}{3})
  \leq \frac{\epsilon'/2}{2}\qty(1+\frac{3}{3}) = \frac{\epsilon'}{2}.
\end{align}
For spectrum-normalized $\hat{H}$, $\norm*{\hat{H}}=\order*{1}$ and thus $\Delta_t = \order*{\epsilon'}$.

A summary of the bounds derived above.
\begin{itemize}
  \item $\displaystyle t_c = \frac{1}{\Gamma} \log\qty(\frac{2}{\Gamma \epsilon'}) $
  \item $\displaystyle \Delta_t = \epsilon' - \frac{W}{3}\epsilon'^2 + \frac{2W^2}{9}\epsilon'^3$
  \item $\displaystyle N_c = \frac{t_c}{\Delta_t} 
  \approx \frac{1}{\Gamma \epsilon'} \log\qty(\frac{2}{\Gamma \epsilon'})$
\end{itemize}

Now we calculate the $\norm{\vec{\alpha}}_{1}$ in the LCU sum $h(\omega+i\Gamma,\hat{H})$.
\begin{align}
     h(\omega+i\Gamma,\hat{H})
  &= \sum_{k=0}^{N_c}
     \Delta_t e^{-\Gamma k\Delta_t} 
     e^{-i[(\hat{H} - \omega)k\Delta_t + \frac{\pi}{2}]}
   = \sum_{k=0}^{N_c} \alpha_k \hat{U}_k, \notag\\
     \norm{\vec{\alpha}}_{1}
  &= \sum_{k=0}^{N_c} \alpha_k
   = \sum_{k=0}^{N_c} \Delta_t e^{-\Gamma k\Delta_t}
   = \Delta_t \frac{1 - e^{-\Gamma(N_c + 1)\Delta_t}}{1 - e^{-\Gamma \Delta_t}} \\
  &< \Delta_t \frac{1}{1 - e^{-\Gamma \Delta_t}}
   = \frac{\Gamma\Delta_t}{1 - e^{-\Gamma \Delta_t}}
     \frac{1}{\Gamma} \\
  &< \frac{\Gamma\epsilon'}{1 - e^{-\Gamma \epsilon'}}
     \frac{1}{\Gamma}
  \quad\left[
   \parbox{3.2cm}{\raggedright
     $\frac{x}{1-e^{-x}}$ monotonically increases with $x$ and $x=\Gamma\Delta_t < \Gamma \epsilon'$}\right
   ] \\
  &= \qty[1 + \frac{\Gamma \epsilon'}{2} + \order{\Gamma^2 \epsilon'^2}] \frac{1}{\Gamma}
   = \frac{1}{\Gamma} + \frac{\epsilon'}{2} + \order{\Gamma \epsilon'^2}
   \approx \frac{1}{\Gamma}. \\
  \therefore \norm{\vec{\alpha}}_{1} 
  &\approx \frac{1}{\Gamma}.
\end{align}
Therefore, the query complexity is $\norm{\vec{\alpha}}_{1} t_c = \frac{1}{\Gamma^2}\log\qty(\frac{2}{\Gamma \epsilon'})$, which only depends on the required precision $\epsilon'$ as $\log(1/\epsilon')$ and in this aspect is similar to the cost of ground state preparation. However, the cost of numerical demonstration on classical computer is proportional to $N_c$ that depends on the required precision $\epsilon'$ as $(1/\epsilon')\log(1/\epsilon')$. For ground state preparation, the total number of LCU terms $N_z$ of the discretized Gaussian integration depends on the required precision $\epsilon'$ as $\log(1/\epsilon')$, so the numerical calculation of the ground state preparation is less expensive than the resolvent calculation on classical computer for the same precision.

\section{Derivation of tightened bounds from Ge \textit{et al.}~\cite{Ge2019}} \label{app:ge_bounds}
For parameter $M$ in Lemma 1 from Ref.~\onlinecite{Ge2019}, we first show that the lower bound of $M$ can be tightened from $M = \Omega\qty(\frac{1}{\Delta(\tau + \delta_E)} \log \frac{1}{\gamma\eta})$ to $M = \Omega\qty(\frac{1}{\Delta(\tau + \delta_E) + (\Delta^2/2)} \log \frac{1}{\gamma\eta})$. Then, we show that by applying the Lemma 1 with this new bound, the required precision of the ground energy in Theorem 1 from Ref.~\onlinecite{Ge2019} can be lowered from $\order{\Delta / \log\frac{1}{\gamma\eta}}$ to $\order{\Delta / \log^{1/2}\frac{1}{\gamma\eta}}$ and the query complexity determined from the parameter $m_0$ can be reduced from $\order{\frac{1}{\Delta}  \log^{3/2}\qty(\frac{1}{\gamma\eta})}$ to $\order{\frac{1}{\Delta} \log\frac{1}{\gamma\eta}}$.

Following the definition of Lemma 1 from Ref.~\onlinecite{Ge2019}, a new Hamiltonian $H$ is defined from the original $\tilde{H}$ as $H = \tilde{H} - (E-\tau) = \tilde{H} - \lambda_0 + (\tau + \delta_E)$, where $E\in [0, \lambda_0]$, $\delta_E = \lambda_0 - E \in [0, \lambda_0]$, and $\tau\in [0,1/2]$. Here $\{ \lambda_0, \ket{\lambda_0} \}$ are the ground energy and ground state of the original $\tilde{H}$ that has a spectrum $\sigma(\tilde{H})\in [\lambda_0, 1] \subseteq [0,1]$ with $\lambda_0 \geq 0$. We denote the first excited energy of $\tilde{H}$ as $\lambda_1 = \lambda_0 + \Delta_s \leq 1$. Here, $E$, $\tau$, $\delta_E$, $\lambda_0$ are not required to be small. Our derivation below also does not require the spectral gap $\Delta_s$ to be small.

Since $\tau + \delta_E \in [0, 3/2]$, we have $\cos^M(\tau + \delta_E) > 0$. For $l > 0$, $\lambda_l - \lambda_0 +\tau + \delta_E \geq \lambda_1 - \lambda_0 + \tau + \delta_E = \Delta_s + \tau + \delta_E > 0$ and $\lambda_l - \lambda_0 + \tau + \delta_E \leq 1 - \lambda_0 + 1/2 + \lambda_0 = 3/2$, so $0 < \cos (\lambda_l - \lambda_0 +\tau + \delta_E) \leq \cos(\Delta_s + \tau + \delta_E)$ for $l>0$. Denote $\tau' = \tau + \delta_E$ for convenience. Note that $\Delta_s + \tau' = \Delta_s + \tau + \delta_E \leq \Delta_s + \tau + \lambda_0 = \lambda_1 + \tau \leq 1 + 1/2 = 3/2$. The new lower bound for $M$ is derived as follows.
\begin{align}
     \norm{\frac{\cos^M (H) \ket*{\lambda_0 ^ \perp}}
       {\cos^M(\tau')}}
  &= \frac{\norm{ \sum_{l > 0} \cos^M (\lambda_l - \lambda_0 +\tau') 
       \qty(c_l \ket*{\lambda_l}) }}
       {\cos^M(\tau')} \\
  &= \frac{ \qty(\sum_{l > 0}\cos^{2M} (\lambda_l - \lambda_0 +\tau') 
       |c_l|^2)^{1/2}}
       {\cos^M(\tau')} \\
  &\leq \frac{ \qty(\cos^{2M} (\Delta_s +\tau') 
       \sum_{l > 0} |c_l|^2)^{1/2}}
       {\cos^M(\tau')} \\
  &= \frac{\sqrt{1 - \gamma^2} \cos^{M} (\Delta_s +\tau')}
       {\cos^M(\tau')}
   < \frac{\cos^{M} (\Delta_s +\tau')}{\cos^M(\tau')} \\
  &= \qty( \frac{\cos\Delta_s \cos\tau' - \sin\Delta_s \sin\tau'}{\cos\tau'} )^M \\
  &= (\cos\Delta_s)^M \qty(1 - \tan\Delta_s \tan\tau')^M \\
  &< (\cos\Delta_s)^M (1 - \tau' \Delta_s)^M \\
  &< e^{-M (\frac{1}{2}\Delta_s^2 + \tau' \Delta_s)}.
\end{align}
In the second to the last step, we used $\tan x > x$ for $x\in (0, \pi/2]$ and $1 - \tan\Delta_s \tan\tau' = \cos(\Delta_s + \tau')/(\cos\Delta_s \cos\tau') > 0$. In the final step, we used $\cos x < e^{-x^2/2}$ for $x\in (0,\pi/2]$, $1-x \leq e^{-x}$ for $x \in \mathbb{R}$, and $(1-x)^M \leq e^{-Mx}$ for $x \leq 1$~\footnote{Both $\tan x > x$ for $x\in (0, \pi/2]$ and $e^{-x} \geq 1-x$ for $x \in \mathbb{R}$ are easy to verify by noticing that the functions $\tan x$ and $e^{-x}$ are concave upward in the respective domains, and $x$ and $1-x$ are the tangent lines for the respective functions in their domains. $\cos x < e^{-x^2/2}$ for $x\in (0,\pi/2]$ can be shown as follows. Integrating both sides of $\tan x > x$ in a domain $[0, \theta]$, where $\theta \in (0, \pi/2)$, we find $-\log\cos\theta > \theta^2/2$. This gives $1/\cos\theta > e^{\theta^2/2}$ and thus $\cos\theta < e^{-\theta^2/2}$ for $\theta \in (0, \pi/2)$. Last, for $\theta=\pi/2$, the inequality can be directly verified.}. %
The last one has a restricted domain due to sign difference between $1-x$ and $(1-x)^M$ for even integer $M$ and $x > 1$. $\tau' \Delta_s < 1$ follows directly from $\tau' \Delta_s < \tan\Delta_s \tan\tau' < 1$. To make
\begin{align}
     \norm{\frac{\cos^M (H) \ket*{\lambda_0 ^ \perp}}
       {\cos^M(\tau')}}
  & < \gamma \eta,
\end{align}
it is sufficient to set $e^{-M (\frac{1}{2}\Delta_s^2 + \tau' \Delta_s)} \leq e^{-M (\frac{1}{2}\Delta^2 + \tau' \Delta)} < \gamma \eta$, and then we find the lower bound for $M$ is
\begin{align}
  M &= \Omega\qty(\frac{1}{\tau' \Delta + (\Delta^2/2)}
       \log \frac{1}{\gamma\eta}).
\end{align}

In proving their Theorem 1 (in which $\Delta_s$ and $\tau'$ are now assumed to be small and $\log \frac{1}{\gamma\eta}$ large), to ensure $\norm{\cos^M(H) \ket{\psi_0}} = \Omega(\gamma)$, Ge \textit{et al.}~\cite{Ge2019} found the following condition must be satisfied: $\tau'^2 M/2 = \order{1}$. For our bound of $M$, this can be achieved by choosing $\tau' = \order{\Delta / \log^{1/2}\frac{1}{\gamma\eta}}$ and $M = \order{\frac{2}{\Delta^2}\log \frac{1}{\gamma\eta}}$. Since $\tau' = \tau + \delta_E$ and $\delta_E$ represents the precision of the ground energy, the required precision of the ground energy is also $\order{\Delta / \log^{1/2}\frac{1}{\gamma\eta}}$. The parameter $m_0$ related to the number of LCU terms and the query complexity can be found by setting the truncation error (by the bound to lower and upper tails of the binomial distribution) $2 e^{-m_0^2/m} = 2 e^{-2m_0^2/M} = \gamma \eta$, which leads to $m_0 = \order{\sqrt{(M/2) \log \frac{2}{\gamma\eta}}} = \order{\frac{1}{\Delta} \log \frac{\sqrt{2}}{\gamma\eta}}$. The constant factor $\sqrt{2}$ is kept in the formula in order to more accurately compare with our algorithm in the numerical demonstration~\footnote{The constant factor $\sqrt{2}$ is derived as follows. 
$(\log \frac{1}{\gamma\eta}
  \log \frac{2}{\gamma\eta})^{1/2}
=[\log \frac{1}{\gamma\eta}
 (\log \frac{1}{\gamma\eta} + \log 2)]^{1/2}
=\log \frac{1}{\gamma\eta}
 [1 + (\log 2) \log^{-1}\frac{1}{\gamma\eta}]^{1/2} 
\approx
 \log \frac{1}{\gamma\eta}
 [1 + (\frac{1}{2}\log 2)\log^{-1}\frac{1}{\gamma\eta}]
=\log \frac{1}{\gamma\eta} + \log\sqrt{2}
=\log \frac{\sqrt{2}}{\gamma\eta}$}.

\bibliography{refGF.bib}

\end{document}